\documentclass[a4paper,fleqn,usenatbib]{mnras}
\usepackage{newtxtext,newtxmath}
\usepackage[T1]{fontenc}
\usepackage{ae,aecompl}
\usepackage{amsmath}
\usepackage{latexsym}
\usepackage{amsfonts}
\usepackage{graphicx}
\usepackage{subfig}
\usepackage{mathrsfs}
\usepackage{epstopdf}
\usepackage{ulem}
\usepackage{bm}
\usepackage{color}
\usepackage{multirow}
\usepackage{url}

\title[Yukawa-type DM/DE interaction]{Testing a quintessence model with Yukawa interaction from cosmological observations and N-body simulations}

\author[R. An et al.]{
Rui An,$^{1,2}$\thanks{E-mail: an\_rui@sjtu.edu.cn}
Andr\'e A. Costa,$^{3,4}$\thanks{E-mail: alencar@if.usp.br}
Linfeng Xiao$^{2}$\thanks{E-mail: hartley@sjtu.edu.cn}
Jiajun Zhang$^{2}$\thanks{E-mail: liamzhang@sjtu.edu.cn}
Bin Wang$^{1,3}$\thanks{E-mail: wang\_b@sjtu.edu.cn}
\\
$^{1}$School of Aeronautics and Astronautics, Shanghai Jiao Tong University, Shanghai 200240, China\\
$^{2}$School of Physics and Astronomy, Shanghai Jiao Tong University, Shanghai 200240, China\\
$^{3}$Center for Gravitation and Cosmology, College of Physical Science and Technology, Yangzhou University, Yangzhou 225009, China\\
$^{4}$Instituto de F\'isica, Universidade de S\~ao Paulo, C.P. 66318, 05315-970, S\~ao Paulo, SP, Brazil\\
}

\def\aap{{ A\&A}}
\def\mnras{{MNRAS}}
\def\apj{{ApJ}}
\def\nat{Nature}
\def\prd{PhRvD}
\def\jcap{JCAP}
\def\apjs{{ApJS}}

\date{Accepted 2019 July 17. Received 2019 July 16; in original form 2019 April 5.}

\pubyear{2019}

\begin{document}
\label{firstpage}
\pagerange{\pageref{firstpage}--\pageref{lastpage}}
\maketitle

\begin{abstract}
We consider a quintessence model with Yukawa interaction between dark energy and dark matter and constrain this model by employing the recent cosmological data including the updated cosmic microwave background (CMB) measurements from Planck 2015,  the weak gravitational lensing measurements from Kilo Degree Survey (KiDS) and redshift-space distortions. We find that an interaction in the dark sector is compatible with observations. The updated Planck data can significantly improve the constraints compared with the previous results from Planck 2013, while the KiDS data has less constraining power than Planck. The Yukawa interaction model is found to be moderately favored by Planck and able to alleviate the discordance between weak lensing measurements and CMB measurements as previously inferred from the standard Lambda cold dark matter model. N-body simulations for Yukawa interaction model is also performed. We find that using the halo density profile is plausible to improve the constraints significantly in the future.
\end{abstract}

\begin{keywords}
dark energy and dark matter -- cosmological observations -- N-body simulations
\end{keywords}

\section{Introduction}

Planck Collaboration has released updated results on the cosmic microwave background anisotropies in 2015 \citep{Adam2016}, which provide the observations on temperature and polarization of the photons from the last scattering surface. The updated data have made significant improvements compared with the previous data in 2013. This allows the derivation of more reliable scientific results and tighter constraints on the cosmological models. 

The standard Lambda cold dark matter ($\Lambda$CDM) model is the most accepted model to explain the cosmic acceleration of our Universe at present. In this model, the driving force of the Universe acceleration is assumed due to the cosmological constant $\Lambda$. Although the $\Lambda$CDM model is proved to be consistent with several observations, it still faces some challenges. Recently, the $\Lambda$CDM model was examined  by employing weak lensing data taken from a $450$-$\text{deg}^2$ observed field of the Kilo Degree Survey (KiDS)~\citep{Hildebrandt2017}, where the cosmic shear is measured from distorted images of distant galaxies which can effectively map a three-dimensional dark matter structure in the late Universe. It was revealed that there exists a ``substantial discordance''  between the KiDS data \citep{Jong2015,Kuijken2015,Hildebrandt2017,Fenech2017} and the Planck 2015 CMB data \citep{Adam2016,Ade2016,Aghanim2016} in the $\Lambda$CDM model at the level of $2.3\sigma$. 

Besides the discordance between weak lensing measurements and the CMB measurements, the standard $\Lambda$CDM model is also challenged by other observations. For example, the value of the Hubble constant which is directly measured by the Hubble Space Telescope (HST) presents about $3\sigma$ tension in comparison with the value inferred from CMB measurements if the $\Lambda$CDM model is considered \citep{Riess2011,Riess2016}. Meanwhile, another evidence against the standard $\Lambda$CDM model has been presented by the Baryon Oscillation Spectroscopic Survey (BOSS) experiment of the Sloan Digital Sky Survey (SDSS) \citep{Delubac2015}, which is based on measurements of the baryon acoustic oscillations (BAO) from the flux correlation functions of the Lyman-$\alpha$ forest with $158, 401$ quasars at high redshifts $(2.1\leq z\leq 3.5)$. Their results indicate a $2.5\sigma$ deviation from the $\Lambda$CDM model in the measurements of the Hubble constant and angular distance at an average redshift $z=2.34$. Recently, the Experiment to Detect the Global Epoch of Reionization Signature (EDGES) reported the detection of an absorption profile in the sky-averaged radio spectrum centered at 78 MHz \citep{Bowman2018}. Their observation indicates a 21 centimeter signal with an amplitude of 0.5 kelvin, which is more than a factor of two greater than the theoretical prediction of the standard $\Lambda$CDM paradigm \citep{Cohen2017}. 

Theoretically, the $\Lambda$CDM model suffers more severe challenges, such as the cosmological constant problem \citep{Weinberg1989}, i.e., the observed value is many orders of magnitude smaller than the prediction from quantum field theory, and the coincidence problem \citep{Chimento2003}, i.e., in the $\Lambda$CDM model it is difficult to explain why the dark energy dominates the evolution in the late Universe and why the Universe is accelerating just now but neither earlier nor later.

Due to these observational and theoretical problems in the standard $\Lambda$CDM model, there are many attempts to find a better model which can solve or alleviate these problems and explain the late time accelerated expansion of our Universe. Considering that dark energy and dark matter are two major components in the Universe, it is natural, in the framework of field theory, to consider that these two dark sectors have some interaction rather than evolve individually. It was argued that an appropriate interaction can provide a mechanism to alleviate the coincidence problem \citep{Amendola2000,Amendola2003,Amendola2006,Pavon2005,Del2008,Boeher2008,Olivares2006,Chen2008}. It can also accommodate an effective dark energy equation of state in the phantom region at the present time \citep{He2009}. However, because of the lack of information on the nature and dynamics of dark energy and dark matter, it is difficult to drive the precise form of the interactions. Many alternative models have been proposed in the literature from phenomenology or field theory \citep{Amendola2000,Amendola2003,Pavon2005,Boeher2008,Olivares2006,Lopez2010,Bertolami2007,Micheletti2009, Costa2017}. For a review on the interaction between dark matter and dark energy, please refer to \citet{Wang2016}. 

In this work we will concentrate on the scenario in which dark matter takes the form of a spin $\frac{1}{2}$ fermionic field and dark energy is described by an evolving and fluctuating scalar field, the quintessence. An interaction between these two components will affect the expansion history of the Universe and the evolution of the density perturbation, changing the growth history of cosmological structure. Consequently, the interaction could be constrained with
observations of the background evolution and the emergence of large scale structure. Following  \cite{Costa2014}, we will consider a Yukawa coupling of the dark energy field to the dark matter, which is renormalizable and has been well studied in cosmology \citep{Farrar2004,Pavan2012}.

The main motivation of this paper is to confront the Yukawa interaction model to the latest cosmological data, including the updated CMB data from Planck 2015, the recent weak gravitational lensing data from KiDS and the redshift-space distortion (RSD) data. We are going to compare the constraints with the previous results from Planck 2013 \citep{Costa2014} and see whether the updated precise data can help to improve the limits on the cosmological parameters. We will also investigate the discordance problem between Planck and KiDS with the Yukawa interaction model and check whether an appropriate interaction can help to alleviate the tension between these two datasets. Meanwhile, the Dark Energy Survey (DES) collaboration has recently published the analyses of its first year of data, based on the two-point statistics of galaxy clustering and weak gravitational lensing \citep{Abbott2018, Abbott2018_2}, which will also be included in our analysis. Moreover, we will examine the effectiveness of tightening the constraints on model parameters by including the complementary RSD observable.

Besides, we will use the ME-Gadget code \citep{me-gadget-prd} to perform several simulations for Yukawa interaction model to investigate the non-linear structure formation. N-body simulations have been established as a standard method to study the non-linear evolution of the large scale structure of the Universe. Because the Yukawa coupling between dark matter and dark energy is absent in the $\Lambda$CDM model, it is important to use a fully self-consistent simulation pipeline to study the non-linear structure formation. Zhang et al. has developed the simulation pipeline suitable for the Yukawa interaction model \citep{me-gadget-prd}.

The paper is organized as follows. In Section II, we describe the Yukawa interaction model with background dynamics equations and linear perturbations. In Section III, we introduce the observational datasets we are going to use. In Section IV, we report the main results by confronting the Yukawa interaction model to cosmological observations. In Section V, we discuss the results from N-body simulations. Finally, we present our conclusions in Section VI.

\section{The Yukawa interaction model}

We consider a model with an interaction between two dark sectors, where dark matter is described by a spin $\frac{1}{2}$ fermionic field and dark energy is described by a canonical scalar field. The action for this model is given by
\begin{equation}
\label{eq.action}
S=\int d^4x\sqrt{-g}\Big\{\frac{1}{2\kappa}R- \frac{1}{2} \partial^{\mu} \phi \partial_{\mu} \phi - V(\phi)- m(\phi) \bar{\psi} \psi + \mathcal{L}_{\mathrm{K}}[ \psi ]\Big\},
\end{equation}
where $g$ is the determinant of the metric, $R$ is the Ricci scalar and $\kappa=8\pi G$ where $G$ represents the gravitational constant. $\phi$ is the scalar field and its potential function $V(\phi)$ can be chosen freely. To be specific, in this paper we will study the exponential form $V(\phi)=Ae^{-\lambda\phi/M_{pl}}$, where $A$ is a normalization constant, $\lambda$ is a dimensionless parameter and $M_{pl} = 1/\sqrt{8 \pi G}$ is the reduced Planck mass. $\psi$ is the fermionic field and $\mathcal{L}_{\mathrm{K}}[ \psi ]$ is the kinetic part of the fermionic Lagrangian. $m(\phi)$ is the effective fermionic mass and its choice represents the coupling to $\phi$. In our model, the function $m(\phi)$ is given by $m(\phi)=M-\zeta\phi$, where $M$ is the fermionic mass and $\zeta$ is the Yukawa coupling constant. This coupling can be treated as an external source in the conservation equations for the dark sectors of the Universe
\begin{align}
\label{eq.Tc}
&\nabla_{\nu} T_{(c) \mu}^{\nu} = -Q_{\mu}, \\
&\nabla_{\nu} T_{(d) \mu}^{\nu} = Q_{\mu},
\label{eq.Td}
\end{align}
where $\nabla_{\nu}$ represents a covariant derivative, $T_{(i) \mu}^{\nu}$ is the stress energy tensor of the `i' component in the Universe, the subscripts `c' denotes dark matter and `d' denotes dark energy.  The source term $Q_{\mu}$ implies that these two components are not conserved, while for the whole system the energy momentum conservation still holds. 

We assume that our Universe is described by a flat Friedmann-Lemaitre-Robertson-Walker (FLRW) metric, in which the line element can be written as
\begin{equation}
\label{eq.metric}
ds^2=-a^2(\eta)d\eta^2+a^2(\eta)\delta_{ij}dx^idx^j,
\end{equation}
where $\eta$ is the conformal time and $a(\eta)$ is the scalar factor of the Universe. For the rest of the paper, a dot will denote the derivative with respect to the conformal time. The zero-component of equations (\ref{eq.Tc}) and (\ref{eq.Td}) provide the conservation equations for the energy densities of the dark sectors
\begin{align}
\label{eq.rhoc}
&\dot{\rho}_c = - 3 \mathcal{H} \rho_c - Q_0, \\ 
&\dot{\rho}_d = - 3 \mathcal{H} \rho_d (1 + \omega) + Q_0,
\label{eq.rhod}
\end{align}
where $\mathcal{H}=\frac{\dot{a}}{a}$ is the Hubble function and $\omega \equiv P_d/\rho_d$ is the dark energy equation of state. Here we treat each component of the dark sector as a fluid with the general stress-energy tensor ${T_{\mu \nu} = (\rho_i + P_i) u_{\mu} u_{\nu} + P_i g_{\mu \nu}}$, where $u_{\mu}=(-a,0,0,0)$ is the fluid 4-velocity. Dark energy is described by a scalar field $\phi$ rolling down a self-interaction potential $V(\phi)$, such that its energy density and pressure can be expressed as
\begin{equation}
\label{scalar}
\rho_d = \frac{\dot{\phi}^2}{2 a^2} + V(\phi), \ P_d = \frac{\dot{\phi}^2}{2 a^2} - V(\phi).
\end{equation}

The external source term $Q_{\mu}$ is related to the effective fermionic mass $m(\phi)$ via the expression
\begin{equation}
\label{eq.Qm}
Q_{\mu} = - \frac{\partial \ln m(\phi)}{\partial \phi} \rho_c \nabla_{\mu} \phi,
\end{equation} 
which gives the coupling term 
\begin{equation}
\label{eq.Q0}
Q_0 = \frac{\zeta}{M - \zeta \phi} \rho_c \dot{\phi} = \frac{r}{1 - r \phi} \rho_c \dot{\phi} ,
\end{equation}
where $r \equiv \frac{\zeta}{M}$. We can rewrite the conservation equations (\ref{eq.rhoc}) and (\ref{eq.rhod}) as 
\begin{align}
\label{dmeq}
&\dot{\rho}_c + 3 \mathcal{H} \rho_c = - \frac{r}{1 - r \phi} \rho_c \dot{\phi}, \\ \label{deeq} 
&\dot{\rho}_d + 3 \mathcal{H} \rho_d (1 + \omega) = \frac{r}{1 - r \phi} \rho_c \dot{\phi}.
\end{align}
To avoid the diverging point at $r\phi=1$, we will stay in the region $r\phi<1$.
The signs of $r$ and $\dot{\phi}$ determine the direction of the energy flow, if they have the same sign the energy flows from dark matter to dark energy while the different sign signaling the opposite. For what concerns the background dynamics, the evolution of the scalar field is described by the modified Klein Gordon equation via
\begin{equation}
\label{eq.phi}
\ddot{\phi}+2\mathcal{H}\dot{\phi}+a^2\frac{\text{d}V}{\text{d}\phi}=a^2\frac{r}{1 - r \phi} \rho_c.
\end{equation}
From the Einstein field equation we can get the Friedmann equation as follows
\begin{equation}
\mathcal{H}^2 = \frac{8 \pi G}{3} a^2 \left( \rho_r + \rho_b + \rho_c + \frac{\dot{\phi}^2}{2 a^2} + V(\phi) \right).
\end{equation}
Here the relativistic component `r' and the baryons `b' are assumed to be uncoupled to the scalar field in this model, hence the evolutions of their energy densities still obey the standard conservation equations
\begin{equation}
\label{eq.rb}
\dot{\rho}_r + 4 \mathcal{H} \rho_r = 0, \ \dot{\rho}_b + 3 \mathcal{H} \rho_b = 0. 
\end{equation}

In the linear theory, equations of the first order perturbations for dark matter can be written as 
\begin{align}
\label{pceq}
&\dot{\delta}_c = - \theta_c - \frac{\dot{h}}{2} - \frac{r}{1 - r \phi} \dot{\varphi} + \frac{r^2}{(1 - r \phi)^2} \dot{\phi} \varphi, \\  
\label{pdeq}
&\dot{\theta_c} = - \mathcal{H} \theta_c + \frac{r}{1 - r \phi} \theta_c \dot{\phi} - k^2 \frac{r}{1 - r \phi} \varphi,
\end{align}
where $\delta_c$ is the perturbed density contrast and $\theta_c=ik_jv_c^j$ is the gradient of velocity field for dark matter. The variable $h$ is the trace part in the synchronous gauge metric perturbation. Perturbation in the scalar field $\varphi \equiv \delta\phi$ evolves according to the perturbed Klein Gordon equation, which can be written as
\begin{align}
\ddot{\varphi} + 2 \mathcal{H} \dot{\varphi} + k^2 \varphi &+ a^2 \frac{\mathrm{d}^2 V}{\mathrm{d} \phi^2} \varphi + \frac{\dot{h} \dot{\phi}}{2} \nonumber \\
&= - a^2 \frac{r^2}{(1 - r \phi)^2} \varphi \rho_c + a^2 \frac{r}{1 - r \phi}\rho_c \delta_c.
\end{align}
For the other components, radiation and baryon, the perturbation equations follow from Boltzmann equations, which are the same as those in the $\Lambda$CDM model.

\section{Cosmological datasets}

We use the latest available results of the CMB measurements from Planck 2015~\citep{Ade2016} to derive constraints for the Yukawa interaction model, which can be directly compared to the previous ones from Planck 2013 \citep{Costa2014}. The updated Planck 2015 data has made significant improvements in reducing the systematic errors and increasing the overall level of confidence. The most notable one is that its residual systematics in polarization maps have been dramatically reduced compared to Planck 2013, and its agreement to WMAP is within a few tenths of a percent on angular scales from the dipole to the first acoustic peak \citep{Adam2016}. These results can make important contributions in the theoretical analyses in cosmology and contain smaller uncertainties compared with those determined in Planck 2013 results.

In \cite{Costa2014} the authors analyse the effects in the CMB and matter power spectrum for the Yukawa interaction model. We can see that the scalar potential parameter $\lambda$ has a small effect on the CMB and matter power spectrum, affecting mainly the low-$l$ CMB power spectrum, while the coupling parameter $r$ not only modifies the CMB spectrum at low-$l$ but also influences the acoustic peaks at large multipoles. These effects allow us to constrain the parameters of such a model through Planck measurements.

In our analysis, we take the low multipole ($l=2-29$) temperature and polarization data, and combine with high multipole ($l\geq 30$) TT, TE and EE CMB data. Also the CMB lensing data will be considered in this work. For the rest of this paper, `Planck13' and `Planck15' denote the datasets including only CMB temperature and polarization spectrum from Planck 2013 and Planck 2015, respectively, `Planck15(+lensing)' denotes Planck15 datasets together with CMB lensing.

In addition to the Planck datasets, we also consider the weak gravitational lensing measurements from the Kilo Degree Survey \citep{Hildebrandt2017}. The KiDS is designed to measure shapes of galaxies with photometric redshifts and it performs a study of weak lensing tomography. The lensing observables are given by the two point shear correlation function $\xi_{\pm}^{ij}$ between two redshifts bins $i$ and $j$ at the angular position $\theta$ on the sky, which can be expressed by the convergence power spectrum $P_{\kappa}^{ij}$ via
\begin{equation}
\label{eq.corP}
\xi_{\pm}^{ij}(\theta)=\frac{1}{2\pi}\int \text{d}llP_{\kappa}^{ij}J_{0,4}(l\theta),
\end{equation}
where $l$ is the angular wave number, and $J_{0,4}(l\theta)$ is the zeroth (for $\xi_+$) or the fourth (for $\xi_-$) order Bessel functions of the first kind. Using the Limber approximation, the convergence power spectrum $P_{\kappa}^{ij}$ can be related to the matter power spectrum $P_{\delta}$ via
\begin{equation}
\label{eq.PP}
P_{\kappa}^{ij}=\int_0^{\chi_H}\text{d}\chi\frac{W_i(\chi)W_j(\chi)}{\chi^2}P_{\delta}\left(\frac{l}{\chi},\chi\right),
\end{equation}
where $\chi$ is the comoving radial distance and $\chi_H$ is the comoving distance evaluated at an infinite redshift. The lensing weighting function $W_i(\chi)$ is given by \citep{Schaefer2008, An2017, An2018}
\begin{equation}
\label{eq.q}
W_i(\chi)=\frac{3a(\chi)^2H(\chi)^2\Omega_m(\chi)}{2c^2}\chi\int_{\chi}^{\chi_H}\text{d}\chi'n_i(\chi')\frac{\chi'-\chi}{\chi'},
\end{equation}
where $\Omega_m=\rho_m/\rho_{\rm{crit}}$ with the critical density $\rho_{\rm{crit}}=3H^2/(8\pi G)$, $c$ is the speed of light, $n_i(\chi)\text{d}\chi$ is the effective number of galaxies in redshift bin $i$ within the range of $\text{d}\chi$ and it is normalized as $\int_0^{\chi_H}n(\chi)\text{d}\chi=1$. We use the modified CAMB code \citep{Lewis2000} to calculate the linear matter power spectrum and the non-linear correction is approximately derived by a halofit model \citep{Takahashi2012}, which is inconsistent with the Yukawa interaction model. Therefore, we use the N-body simulation, which is modified to be consistent with the interacting dark matter and dark energy model \citep{me-gadget-prd}, to test whether the halofit model can provide reasonable calculation of the non-linear matter power spectrum for the Yukawa interaction model. The N-body simulation results are introduced in Section V.

The KiDS datasets consist of four tomographic redshift bins between $z=0.1$ to $z=0.9$ with equal widths $\Delta z=0.2$, and nine angular bins with central values at $\theta$ = [$0.7134'$, $1.452'$, $2.956'$, $6.017'$, $12.25'$, $24.93'$, $50.75'$, $103.3'$, $210.3'$]. For each tomographic redshift pair ($ij$), the measurements cover seven angular bins smaller than $72'$ for $\xi_+^{ij}$ and six angular bins larger than $4.2'$ for $\xi_-^{ij}$, which means that the last two angular bins are marked out for $\xi_+^{ij}$ and the first three bins are marked out for $\xi_-^{ij}$. This equates to a total of $130$ angular band powers in this datasets \citep{Hildebrandt2017,Joudaki2017}. 

Cosmic shear has also been recently measured using DES data \citep{Abbott2018,Troxel2018}. The DES is a five year observing program using the 570 megapixel Dark Energy Camera \citep{Flaughe2015} on the Blanco telescope at the Cerro Tololo Inter-American Observatory, to image the South Galactic Cap in the $griz$Y filters. DES collaboration has presented the cosmological results from a combined analysis of galaxy clustering and weak gravitational lensing, using $1321\text{deg}^2$ of $griz$ imaging data taken during its first year of observations, where three sets of two-point correlation functions are included: (i) Galaxy clustering: the auto-correlation of lens galaxy positions in each redshift bin $\omega(\theta)$, (ii) Cosmic shear: the auto-correlation of source galaxy shapes within and between the source redshift bins $\xi_{\pm}(\theta)$, (iii) Galaxy-galaxy lensing: the mean tangential ellipticity of source galaxy shapes around lens galaxy positions for each pair of redshift bins $\gamma_t(\theta)$  \citep{Abbott2018}. In our analysis, we make use of all three correlation function measurements to test the Yukawa interaction model. Although in \cite{Planck2018}, the authors showed some consistency between DES and Planck 2018 data, after some more careful photo-$z$ calibration, DES result was found more consistent with KiDS and thus showed more tension with Planck 2018 data \citep{Joudaki2019}. In this work, we will examine whether the Yukawa interacting dark energy model can alleviate the tension between Planck and weak lensing datasets.

Another observable employed in this work, RSD, is an important probe to investigate the growth of large scale structure, and it is considered as a powerful complementary observation to obtain tight constrains on cosmological parameters, and also break the possible degeneracy in diverse cosmologies. In the past few years, a lot of measurements on RSD have been reported \citep{Song2009,Samushia2012,Blake2011,Tojeiro2012,Reid2012,Beutler2012,Hudson2012,delaTorre2013,Feix2015}. We list the low-$z$ measurement on $f\sigma_8$ from previous work in Table \ref{RSD}. Note that the data at $z=0.02$ \citep{Hudson2012} is not a result from RSD observation but one inferred from the peculiar velocity obtained during distance measurement.

\begin{figure}
\includegraphics[width=0.45\textwidth]{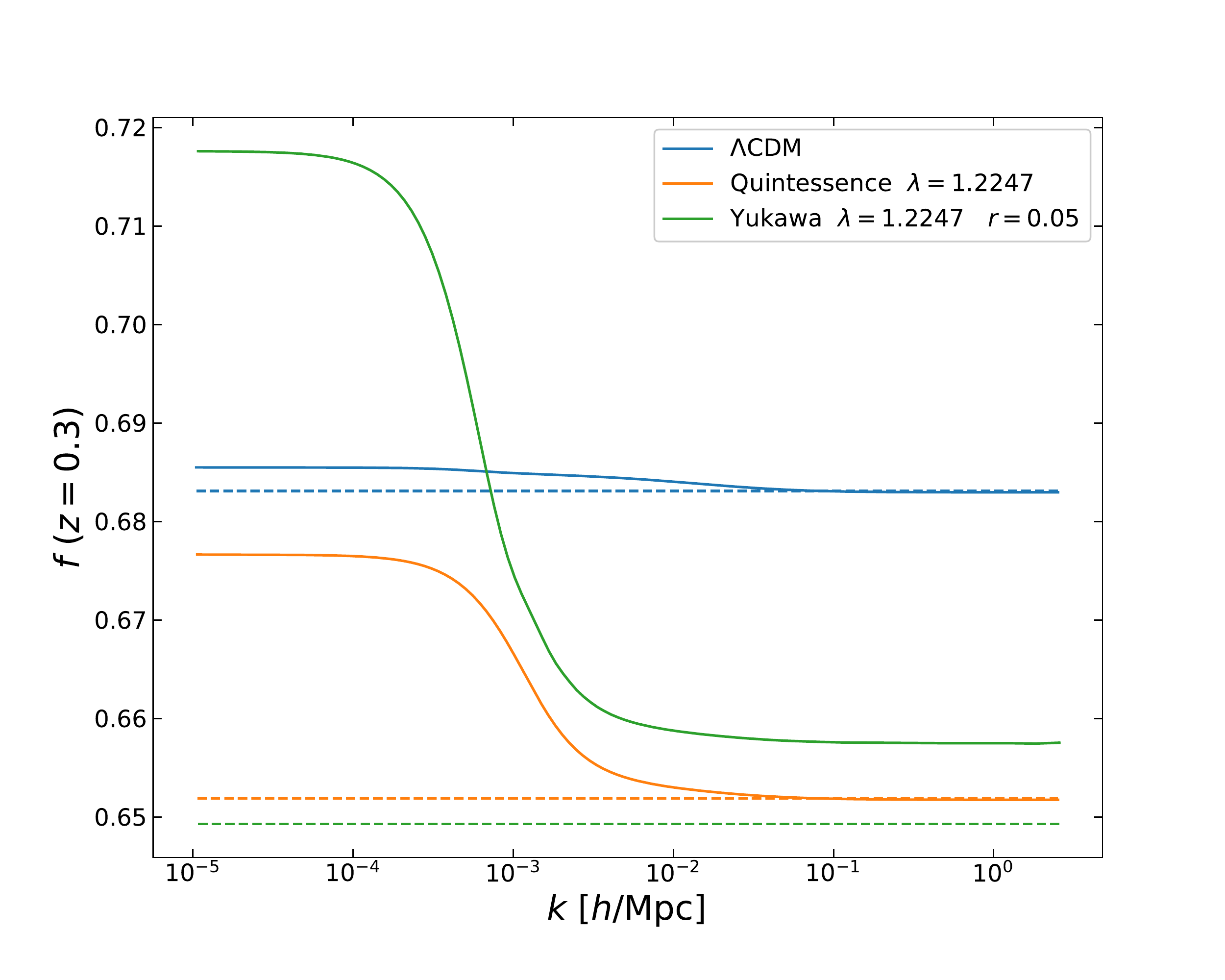}
\caption{The growth rate $f$ as a function of wavenumber $k$ at redshift $z=0.3$. The solid and dashed lines correspond to $f \equiv \frac{\text{dln}\delta_m}{\text{dln}a}$ and $f \equiv [\sigma_8^{(vd)}(z)]^2/[\sigma_8^{(dd)}(z)]^2$, respectively.}
\label{fig.f}
\end{figure}

The growth rate $f$ is considered to be model independent, which is defined as the logarithmic derivative of total matter density perturbation with respect to the logarithm of scale factor 
\begin{equation}\label{define_f} 
f \equiv \frac{\text{dln}\delta_m}{\text{dln}a} = \frac{\mathcal{H}^{-1}}{\delta_m}\dot{\delta}_m,
\end{equation}
where $\delta_m = (\rho_c\delta_c+\rho_b\delta_b)/\rho_m$. From Fig. \ref{fig.f} we can see that another definition \citep{Ade2016}
\begin{equation}\label{define_f2} 
f \equiv [\sigma_8^{(vd)}(z)]^2/[\sigma_8^{(dd)}(z)]^2,
\end{equation}
where $\sigma_8^{(dd)}$ is the variance of density field smoothed within $8h^{-1}$Mpc and $\sigma_8^{(vd)}$ is the same scale smoothed velocity-density correlation, fits well with Eq. (\ref{define_f}) for the $\Lambda$CDM model, as well as the quintessence model on small scale. We present a brief derivation below, to explain why these two definitions are equivalent in the standard $\Lambda$CDM model.

In linear perturbation theory, assuming baryon tracing cold dark matter, we have
\begin{equation}\label{eq:continuity}
\delta_m' = -kv_{m} - 3\phi',
\end{equation}
where $k$ is the wavenumber, $v_{m}$ is the peculiar velocity of matter, $\phi$ is the gravitational potential in Newtonian gauge. The contribution from $3\phi'$ can be neglected on small scales. By using Eq. (\ref{define_f}), we can get the relation between between $v_{m}$ and $\delta_m$,
\begin{equation}\label{eq:kvN}
v_{m} = -\frac{\delta_m' }{k}= -\mathcal{H}f\frac{\delta_m}{k}.
\end{equation}
While from observations, what we really obtain is $v=-\nabla \cdot \textbf{v}_{m}/\mathcal{H}$. Correlating $\delta_m$ and $v$ in Fourier space where $v=-kv_{m}/\mathcal{H}$, we have
\begin{equation}\label{eq:corre}
<\delta_mv> = f<\delta_m\delta_m>,
\end{equation}
which is equivalent to $P_{\delta_mv}(k) = fP_{\delta_m\delta_m}(k)$. After that, we calculate $\sigma_8^{(vd)}$ and $\sigma_8^{(dd)}$ through the convention
\begin{align}\label{eq:s8}
[\sigma_8^{(vd)}]^2 &= \int P_{\delta_mv}(k)W^2(kR)k^2dk,\\
[\sigma_8^{(dd)}]^2 &= \int P_{\delta_m\delta_m}(k)W^2(kR)k^2dk,
\end{align}
where $W(kR)$ is a top-hat filter on the scale $R=8h^{-1}$Mpc. Since $f$ is $k$-independent, we obtain
\begin{equation}\label{eq:def2}
[\sigma_8^{(vd)}]^2 = f[\sigma_8^{(dd)}]^2,
\end{equation}
which proves that these two definitions (\ref{define_f}) and (\ref{define_f2}) are equivalent in the standard $\Lambda$CDM model.

However, for the Yukawa interaction model, as shown in Fig. \ref{fig.f}, the second definition (\ref{define_f2}) does not fit Eq. (\ref{define_f}), which indicates that this definition is not generally valid for the interaction models \citep{Costa2017,Marcondes2016,Costa2019}. Starting from Eq. (\ref{dmeq}) and Eq. (\ref{pceq}), we obtain the form of Eq. (\ref{define_f}) under Yukawa interaction in the synchronous gauge
\begin{equation}\label{Yukawa_f}
  f = \frac{\mathcal{H}^{-1}}{\delta_m}\Big \{ -\theta_m-\frac{\dot{h}}{2}+\frac{\rho_c}{\rho_m}\Big [\frac{r}{1 - r \phi}(\dot{\phi}\delta_m-\dot{\phi}\delta_c-\dot{\varphi})+\frac{r^2}{(1 - r \phi)^2} \dot{\phi} \varphi\Big]\Big \}
\end{equation}
where $\theta_m = (\rho_c\theta_c+\rho_b\theta_b)/\rho_m$. In addition to the different cosmic evolution from the $\Lambda$CDM model, Yukawa interaction also induces an extra term that is proportional to $\rho_c/\rho_m$ on the right side of Eq. (\ref{Yukawa_f}).  Both of these effects account for the misalignment between green solid and dashed lines in Fig. \ref{fig.f}. This expression of $f$ intrinsically describes the growth rate in the context of Yukawa interaction.

\begin{table}
        \caption{RSD data} \centering \label{RSD}
	\begin{tabular}{ccccc}
		\hline
		\multicolumn{1}{c}{\ \ \ \ \ z\ \ \ \ \ } & \multicolumn{1}{c}{f$\sigma_8(z)$} & \multicolumn{1}{c}{Reference}\ \  \\
		\hline
		0.02	 & 0.360 $\pm$ 0.040 & \citep{Hudson2012} \\
		0.067 & 	0.423 $\pm$ 0.055 & \citep{Beutler2012} \\
		0.10	 & 0.37 $\pm$	0.13 & \citep{Feix2015} \\
		0.17	 & 0.51 $\pm$	0.06  & \citep{Song2009} \\
		0.22	 & 0.42 $\pm$	0.07 & \citep{Blake2011} \\
		0.25	 & 0.3512 $\pm$ 0.0583 & \citep{Samushia2012} \\
		0.30	 & 0.407 $\pm$ 0.055 & \citep{Tojeiro2012} \\
		0.35	 & 0.440 $\pm$ 0.050 & \citep{Song2009} \\
		0.37	 & 0.4602 $\pm$ 0.0378 & \citep{Samushia2012} \\
		0.40	 & 0.419 $\pm$ 0.041 & \citep{Tojeiro2012} \\
		0.41	 & 0.45 $\pm$	0.04 & \citep{Blake2011} \\
		0.50	 & 0.427 $\pm$ 0.043 & \citep{Tojeiro2012} \\
		0.57	 & 0.427 $\pm$ 0.066 & \citep{Reid2012} \\
		0.6	 & 0.43 $\pm$	0.04 & \citep{Blake2011} \\
		0.6	& 0.433 $\pm$ 0.067 & \citep{Tojeiro2012} \\
		0.77	 & 0.490 $\pm$ 0.180 & \citep{Song2009} \\
		0.78	 & 0.38 $\pm$	0.04 & \citep{Blake2011} \\
		0.80	 & 0.47 $\pm$	0.08 & \citep{delaTorre2013} \\
		\hline
	\end{tabular}
\end{table}

In $\Lambda$CDM model, the galaxy continuity equation $\theta_G = -\mathcal{H} \beta \delta_G - \frac{\dot{h}}{2}$ which characterizes the coherent motion of galaxies, is built upon the matter continuity equation $\theta_m = -\mathcal{H} f \delta_m - \frac{\dot{h}}{2}$ via the assumption that galaxies trace the matter field according to $\delta_G=b\delta_m$ and $\theta_G=\theta_m$. The measurements of RSD parameter $\beta=f/b$ are based on its correspondence with velocity divergence as established by the continuity equation. However in the interaction models, this continuity equation does not hold anymore \citep{Costa2017,Marcondes2016,Costa2019}, we need to find the correct quantity that corresponds to the velocity field.

In Yukawa interaction model, the continuity equation (\ref{pceq}) for dark matter in the first order perturbations includes an extra interaction term. For baryon, we still have $\dot{\delta}_b = - \theta_b - \frac{\dot{h}}{2}$. Combining these two equations, we get the continuity equation for total matter altered by Yukawa interaction
\begin{equation}
\theta_m = -\mathcal{H} \tilde{f} \delta_m - \frac{\dot{h}}{2}
\end{equation}
where
\begin{equation}\label{Yukawa_f_eff}
\tilde{f} = \frac{\text{dln}\delta_m}{\text{dln}a} - \frac{\mathcal{H}^{-1}}{\delta_m}\frac{\rho_c}{\rho_m}\Big[\frac{r}{1 - r \phi}(\dot{\phi}\delta_m-\dot{\phi}\delta_c-\dot{\varphi})+\frac{r^2}{(1 - r \phi)^2} \dot{\phi} \varphi \Big]
\end{equation}
is the growth rate which makes the continuity equation compatible with the RSD measurements for the Yukawa interaction model. We find that the extra term on the right side of this equation can completely cancel the extra term in equation (\ref{Yukawa_f}).

\begin{figure}
	\subfloat[$r=0$]{
    \includegraphics[width=0.45\textwidth]{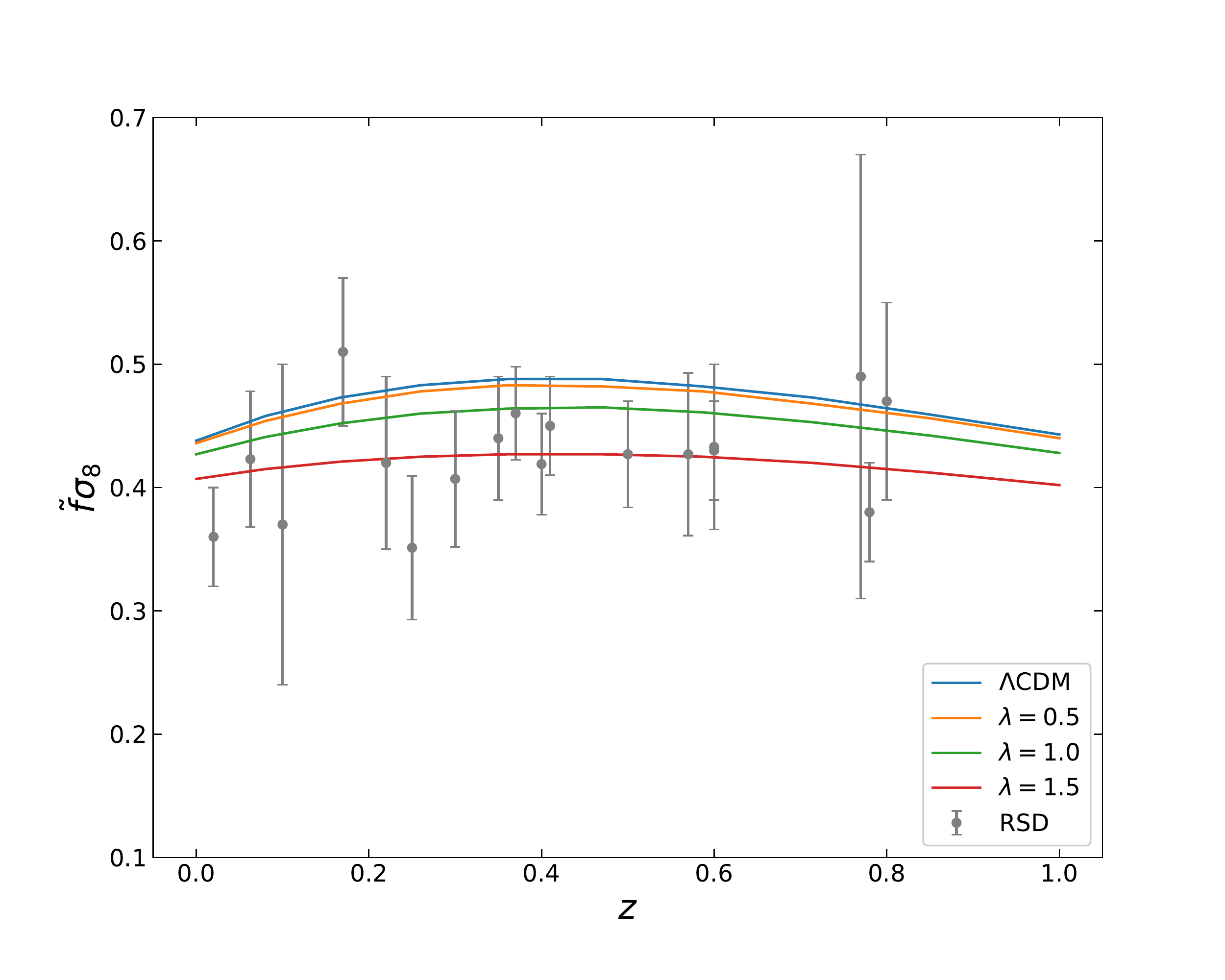}\label{fs8_r}}\\
    \subfloat[$\lambda=1.2247$]{
    \includegraphics[width=0.45\textwidth]{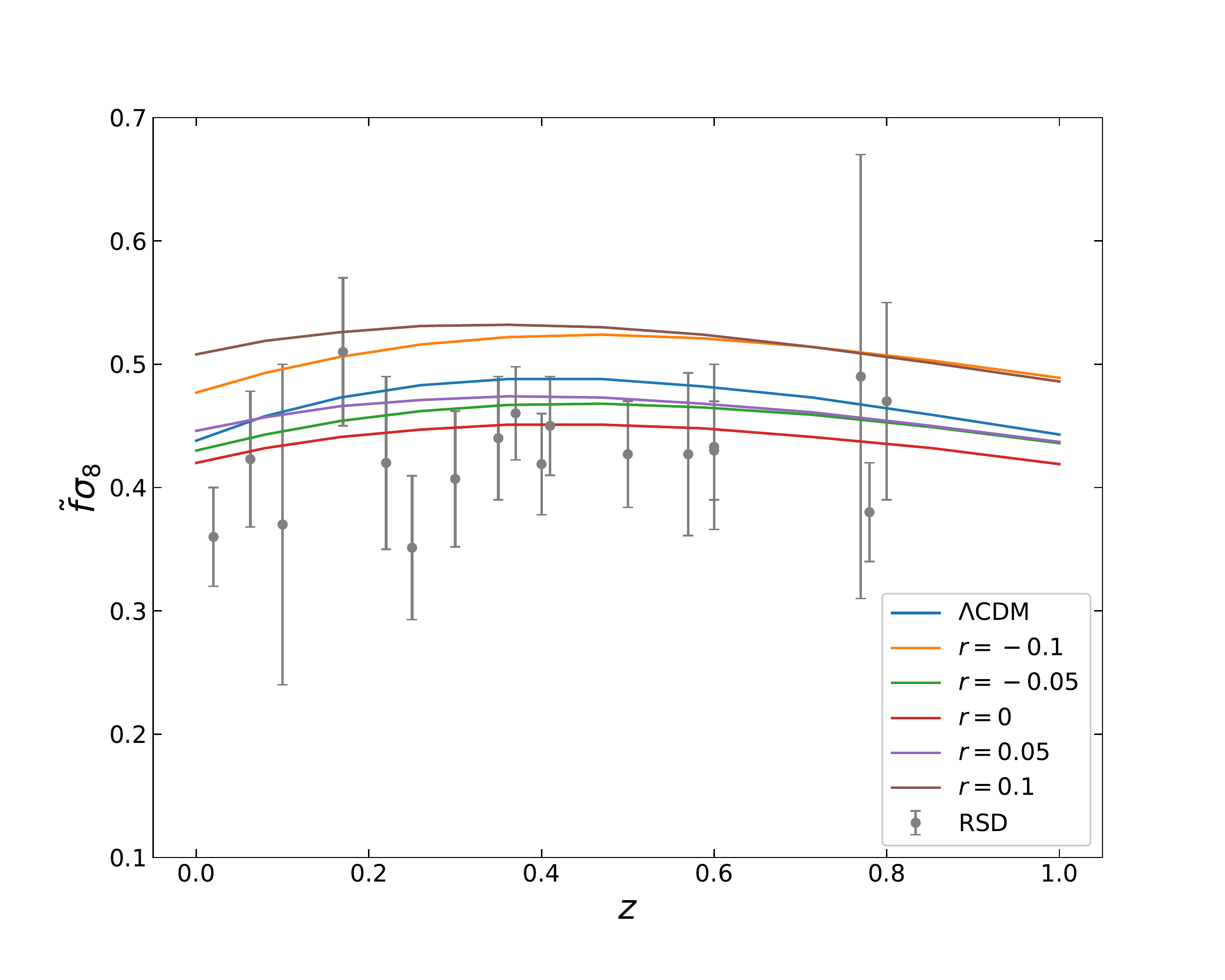}\label{fs8_L}}
	\caption{Evolutions of $\tilde{f}\sigma_8$ as a function of redshift $z$ at $k = 0.1h/\text{Mpc}$ in the $\Lambda$CDM and Yukawa interaction models. The error bars correspond to the RSD data listed in Table. \ref{RSD}.}
	\label{fig.fs8}
\end{figure}

In Fig. \ref{fig.fs8}, we show the evolutions of $\tilde{f}\sigma_8$ at $k = 0.1h/\text{Mpc}$ ($k \gg \mathcal{H}$) in $\Lambda$CDM and Yukawa interaction models. Here the $\Lambda$CDM model is implemented with the best-fitted parameters from Planck 2015 results \citep{Ade2016} where $\Omega_bh^{2}=0.02225$, $\Omega_ch^{2}=0.1198$, $100\theta_{MC}=1.04077$, $\text{ln}(10^{10}A_s)=3.094$, $n_s=0.9645$ and $H_{0}=67.27$km/s/Mpc. We can see that the low-$z$ evolution of $\tilde{f}\sigma_8$ in $\Lambda$CDM model is slightly higher than the observational data, which implies some freedom for non-standard cosmological models to alleviate this discordance. We first present the results for non-interacting models ($r=0$) with different scalar potential parameter $\lambda$. As shown in Fig. \ref{fs8_r}, $\tilde{f}\sigma_8$ is depressed with the increase of $\lambda$, which is compatible with the results in Fig. 1 of \cite{Costa2014}. For the quintessence model without interaction, larger $\lambda$ leads to higher dark energy density at late universe, this in turn will bring more suppression to the growth of overdensity. Then we fix $\lambda$ and extend our discussion to interaction models with varying coupling parameter $r$. In Fig. \ref{fs8_L}, we observe that the Yukawa interaction prefers to increase the values of $\tilde{f}\sigma_8$ comparing to the non-interaction model $r=0$ (red line), and this phenomenon manifests some symmetry around the axis of $r=0$. But it is not absolutely symmetric, we can see that the misalignment appears when $z\lesssim0.6$. From above discussions, we conclude that the Yukawa interaction model with large $\lambda$ and small $r\sim 0$ is more compatible with the RSD measurements at late universe.

Note that \cite{Costa2014} already put constraints on this Yukawa interaction model by using the CMB measurements from the Planck satellite together with BAO, SNIa and H0 data, and detailedly investigate the evidence induced by these extra low redshift measurements. So these three datasets will not be included in this work.

In order to test the Yukawa interaction model, we implement the background and linear density perturbation equations as described in the previous section into the CAMB code \citep{Lewis2000} and, then, use a modified CosmoMC code package \citep{Lewis2002, Lewis2013} which has already integrated the weak lensing module to estimate the parameters that best describe the observational data \citep{Joudaki2017}. For the MCMC runs, we fix the effective number of neutrino species to $N_{\rm{eff}} = 3.046$, the sum of neutrino masses to $\Sigma m_{\nu}= 0.06\,\rm{eV}$ and the helium abundance to $Y_p=0.24$. The convergence criterion is set to $R-1 = 0.03$ where $R$ is the Gelman--Rubin threshold~\citep{Gelman1992}.

\section{Fitting Results} 

We consider a Yukawa interaction between two dark sectors and constrain this model by employing the cosmological datasets introduced in the previous section. We first report the results by using the CMB data from Planck and the weak gravitational lensing data from KiDS. In our numerical analysis we have let the coupling parameter $r$ and the scalar potential parameter $\lambda$ to vary freely. The flat priors on the cosmological parameters are chosen the same as the ones in \cite{Costa2014}, listed in Table \ref{tab.prior}, so that the comparisons can be carried out by using the Planck13 results obtained in \cite{Costa2014} and our new results from Planck15, and also the concordance problem can be examined between the Planck datasets and the KiDS datasets. 

\begin{table}
\centering
\caption{Priors on cosmological parameters \label{tab.prior}}
\begin{tabular}{c|c|c|c|c}
\hline
\ \ \ \ \ Parameter\ \ \ \ \  & \multicolumn{4}{|c}{\ \ \ \ \ \ \ \ Prior\ \ \ \ \ \ \ \ }\\
\hline
$\Omega_bh^2$ & \multicolumn{4}{|c}{[0.005,0.1]}\\
\hline
$\Omega_ch^2$ & \multicolumn{4}{|c}{[0.001,0.99]}\\
\hline
$100\theta$ & \multicolumn{4}{|c}{[0.5,10]}\\
\hline
$\tau$ & \multicolumn{4}{|c}{[0.01,0.8]}\\
\hline
$n_s$ & \multicolumn{4}{|c}{[0.9,1.1]}\\
\hline
$\text{log}(10^{10}A_s)$ & \multicolumn{4}{|c}{[2.7,4]}\\
\hline
$\lambda$ & \multicolumn{4}{|c}{[0.1,1.5]}\\
\hline
$r=\frac{\zeta}{M}$ & \multicolumn{4}{|c}{[-0.1,0.1]}\\
\hline
\end{tabular}
\end{table}

\begin{table*}
\centering
\caption{Best fit values and $68\%$ confidence levels for the cosmological parameters}
\label{tab.pk}
\begin{tabular}{ccccccccc}
\hline
    & \multicolumn{2}{c}{Planck13} & \multicolumn{2}{c}{Planck15} & \multicolumn{2}{c}{Planck15(+lensing)} & \multicolumn{2}{c}{KiDS} \\
    \cline{2-9}
    Parameter & Best fit  & 68\% limits & Best fit  & 68\% limits & Best fit  & 68\% limits & Best fit  & 68\% limits \\
    \hline
$\Omega_bh^2$ & 0.02186 & $0.02195^{+0.000279}_{-0.00028}$ & 0.02203 & $0.0222^{+0.000157}_{-0.000157}$ & 0.02225 & $0.0222^{+0.000156}_{-0.000156}$ & 0.01982 & $0.03785^{+0.0103}_{-0.0287}$\\
$\Omega_ch^2$ & 0.1159 & $0.1171^{+0.00477}_{-0.00315}$ & 0.1201 & $0.1181^{+0.00305}_{-0.00157}$ & 0.1196 & $0.1174^{+0.00331}_{-0.00161}$ & 0.1126 & $0.1239^{+0.0179}_{-0.0281}$\\
$100\theta_{MC}$ & 1.041 & $1.041^{+0.000651}_{-0.000645}$ & 1.041 & $1.041^{+0.000336}_{-0.000338}$ & 1.041 & $1.041^{+0.000335}_{-0.000334}$ & 1.152 & $1.077^{+0.0489}_{-0.064}$\\
$\tau$ & 0.08589 & $0.08879^{+0.0125}_{-0.0139}$ & 0.0814 & $0.08287^{+0.0172}_{-0.017}$ & 0.06774 & $0.06663^{+0.0133}_{-0.0133}$ & 0.23 & $0.3224^{+0.101}_{-0.312}$\\
$n_s$ & 0.9589 & $0.959^{+0.0075}_{-0.00753}$ & 0.9632 & $0.9644^{+0.00488}_{-0.00485}$ & 0.9663 & $0.9653^{+0.00478}_{-0.00479}$ & 0.9434 & $1.001^{+0.0993}_{-0.101}$\\
$\text{log}(10^{10}A_s)$ & 3.084 & $3.086^{+0.0246}_{-0.0249}$ & 3.1 & $3.102^{+0.0334}_{-0.0331}$ & 3.068 & $3.067^{+0.0246}_{-0.0243}$ & 3.555 & $3.307^{+0.693}_{-0.607}$\\
$\lambda$ & 0.5627 & $0.7497^{+0.75}_{-0.65}$ & 0.2902 & $0.6858^{+0.231}_{-0.676}$ & 1.098 & $0.7084^{+0.792}_{-0.698}$ & 0.9991 & $0.8493^{+0.651}_{-0.256}$\\
$r$ & -0.06695 & $-0.009795^{+0.046}_{-0.0613}$ & -0.02669 & $-0.01074^{+0.0424}_{-0.0426}$ & -0.01847 & $-0.009974^{+0.0463}_{-0.0469}$ & 0.07272 & $0.004417^{+0.0956}_{-0.104}$\\
\hline
$\Omega_\Lambda$ & 0.7175 & $0.6882^{+0.03}_{-0.037}$ & 0.6827 & $0.6835^{+0.0224}_{-0.0268}$ & 0.6608 & $0.6873^{+0.0241}_{-0.0285}$ & 0.8547 & $0.7895^{+0.0559}_{-0.0437}$\\
$\Omega_m$ & 0.2825 & $0.3118^{+0.037}_{-0.03}$ & 0.3173 & $0.3165^{+0.0268}_{-0.0224}$ & 0.3392 & $0.3127^{+0.0285}_{-0.0241}$ & 0.1453 & $0.2105^{+0.0437}_{-0.0559}$\\
$z_{\rm{re}}$ & 10.71 & $10.94^{+1.08}_{-1.08}$ & 10.37 & $10.36^{+1.62}_{-1.37}$ & 9.026 & $8.86^{+1.32}_{-1.16}$ & 23.11 & $20.07^{+8.68}_{-12.}$\\
$H_0$ & 69.99 & $67.16^{+2.41}_{-3.3}$ & 67.09 & $66.89^{+2.09}_{-2.3}$ & 64.8 & $67.13^{+2.14}_{-2.59}$ & 95.67 & $88.45^{+11.6}_{-3.93}$\\
${\rm{Age}}/{\rm{Gyr}}$ & 13.65 & $13.75^{+0.157}_{-0.0744}$ & 13.81 & $13.76^{+0.112}_{-0.037}$ & 13.83 & $13.75^{+0.125}_{-0.0435}$ & 11.19 & $11.55^{+0.853}_{-1.37}$\\
\hline
\end{tabular}
\end{table*}

\begin{figure*}
 \includegraphics[scale=0.36]{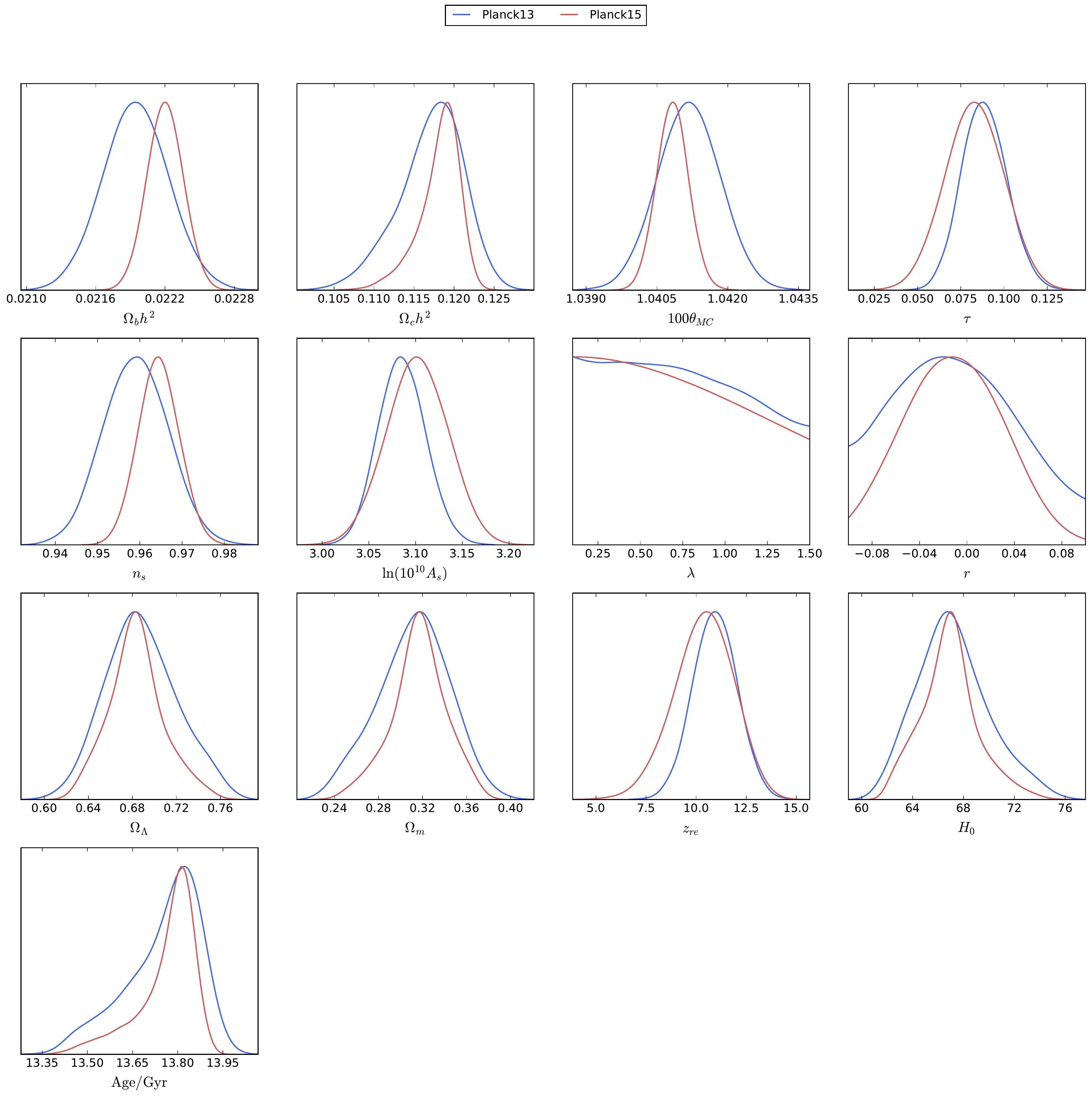}
\caption{1D distributions for the cosmological parameters using Planck data.}
\label{fig.1D_p}
\end{figure*}

\begin{figure*}
 \includegraphics[scale=0.5]{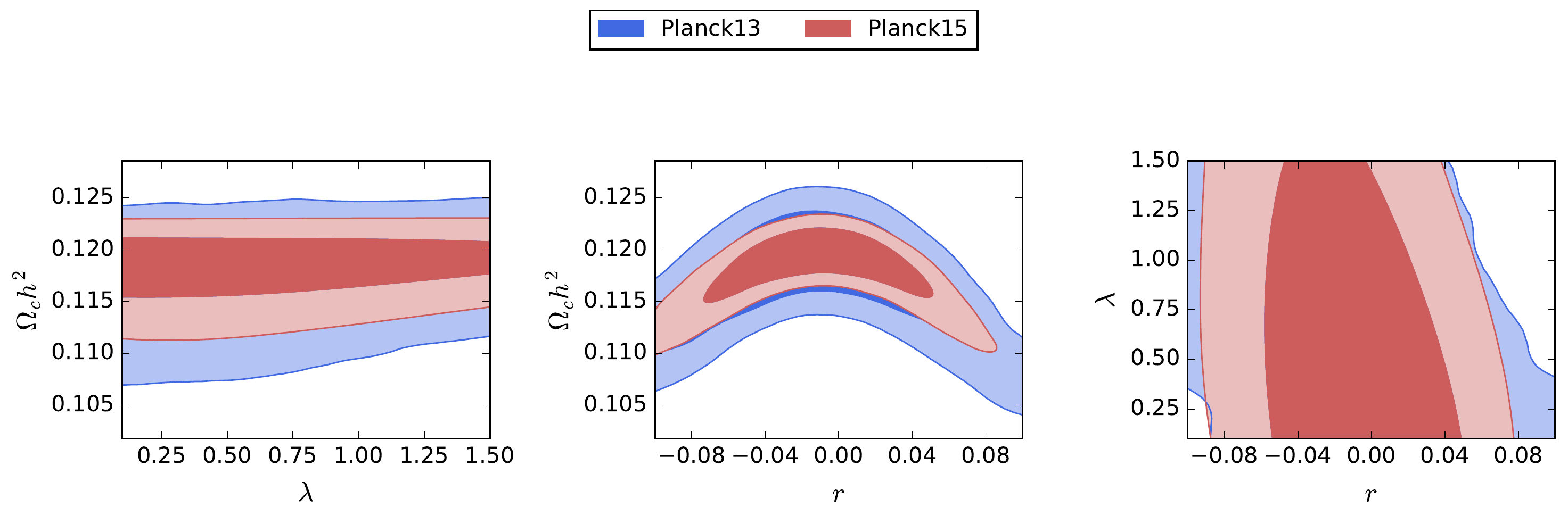}
\caption{2D distributions for selected parameters using Planck data.}
\label{fig.2D_p}
\end{figure*}

\begin{figure*}
 \includegraphics[scale=0.36]{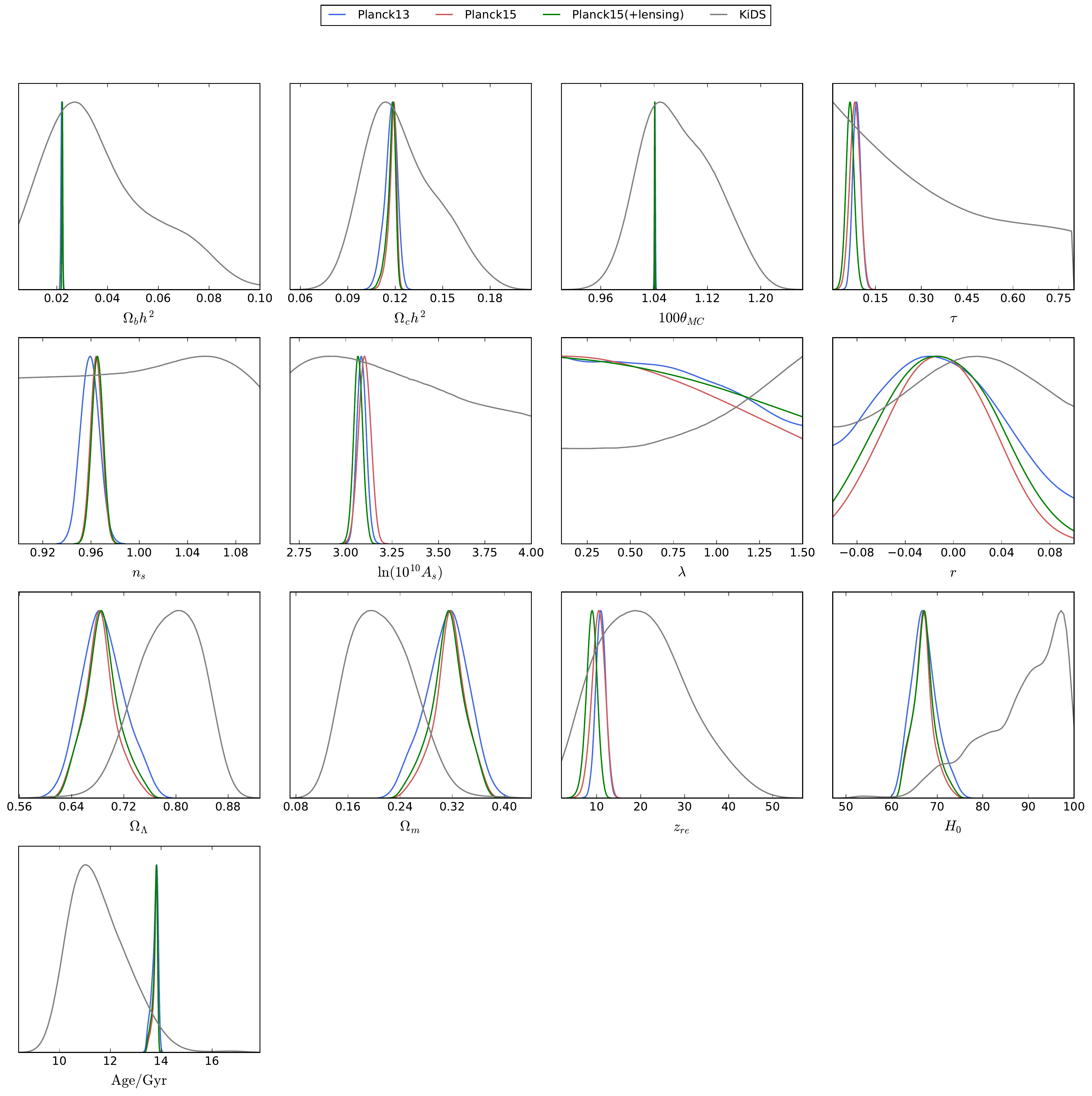}
\caption{1D distributions for the cosmological parameters using Planck and KiDS data.}
\label{fig.1D_plk}
\end{figure*}

The constraints on the parameters and the best fit values are reported in Table \ref{tab.pk}, where we also include the previous results in \cite{Costa2014} for comparison. Here the mean values and $68\%$ limits are obtained from the marginalized distribution, and the best fit values are determined by the maximum N-dimensional likelihood of the samples. Fig. \ref{fig.1D_p} shows the 1D marginalized posterior distributions by using Planck datasets, and the 2D distributions for some parameters of interest are plotted in Fig. \ref{fig.2D_p}. We employ a criteria--Figure of Merit (FoM)--to evaluate the constraining ability of the updated Planck dataset. FoM can be defined in different ways, as long as its value can reflect how tightly or loosely the data constrain parameters. Here for the convenience of our analysis, we adopt the simple definition $\text{FoM}=S_{\text{Planck15}}/S_{\text{Planck13}}$ where $S$ represents the area of $1\sigma$ region. The values of FoM for $\Omega_{c}h^2-\lambda$, $\Omega_{c}h^2-r$ and $\lambda-r$ planes are 1.582, 2.378 and 1.274 respectively, which indicates that Planck15 data can produce a significant improvement in the constraints compared with the previous results by using Planck13 data. We can see that the $1\sigma$ range for the coupling parameter $r$ is much smaller and the best fit value of it becomes less negative by using new Planck data. For the scalar potential parameter $\lambda$, the new Planck data has improved the $1\sigma$ range but it is still not enough to constrain this parameter as shown in Fig. \ref{fig.1D_p}. From Fig. \ref{fig.2D_p} we find that the degeneracies between the parameters $\Omega_ch^2$, $\lambda$ and $r$ do not have any significant difference from the previous results obtained from Planck13. 

\begin{figure}
    \includegraphics[width=0.23\textwidth]{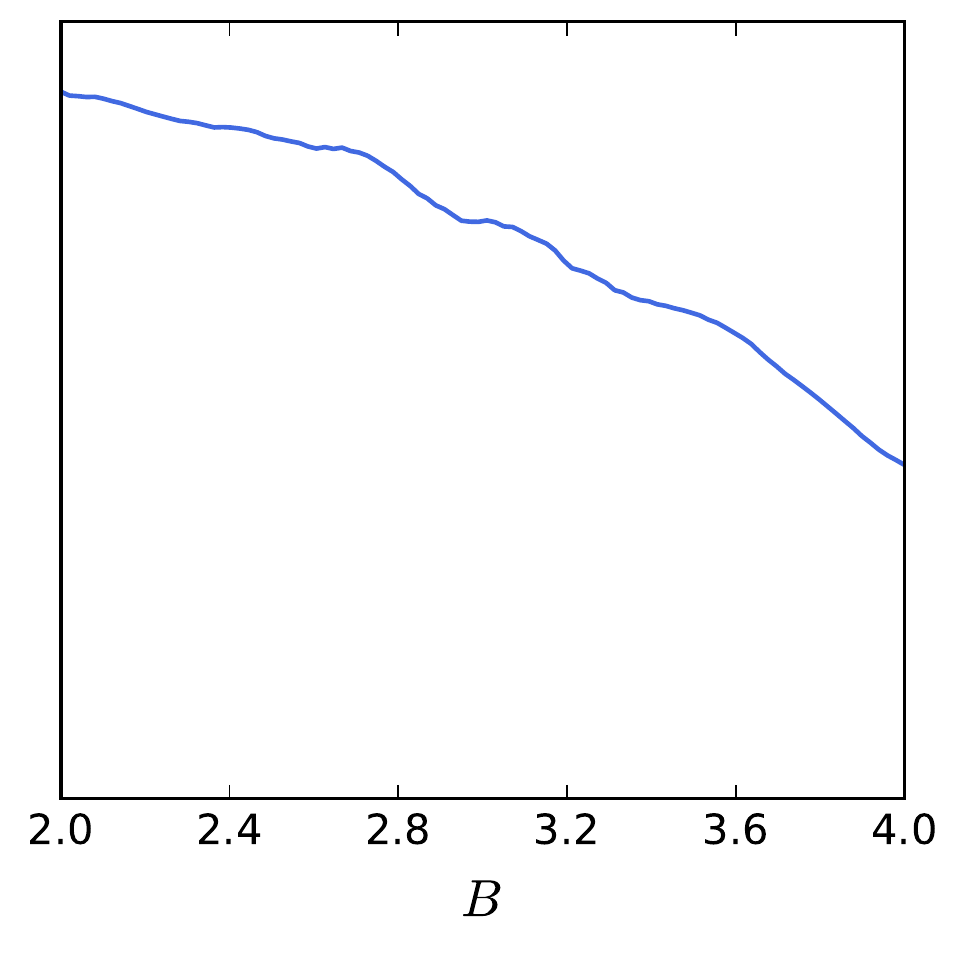}
    \includegraphics[width=0.23\textwidth]{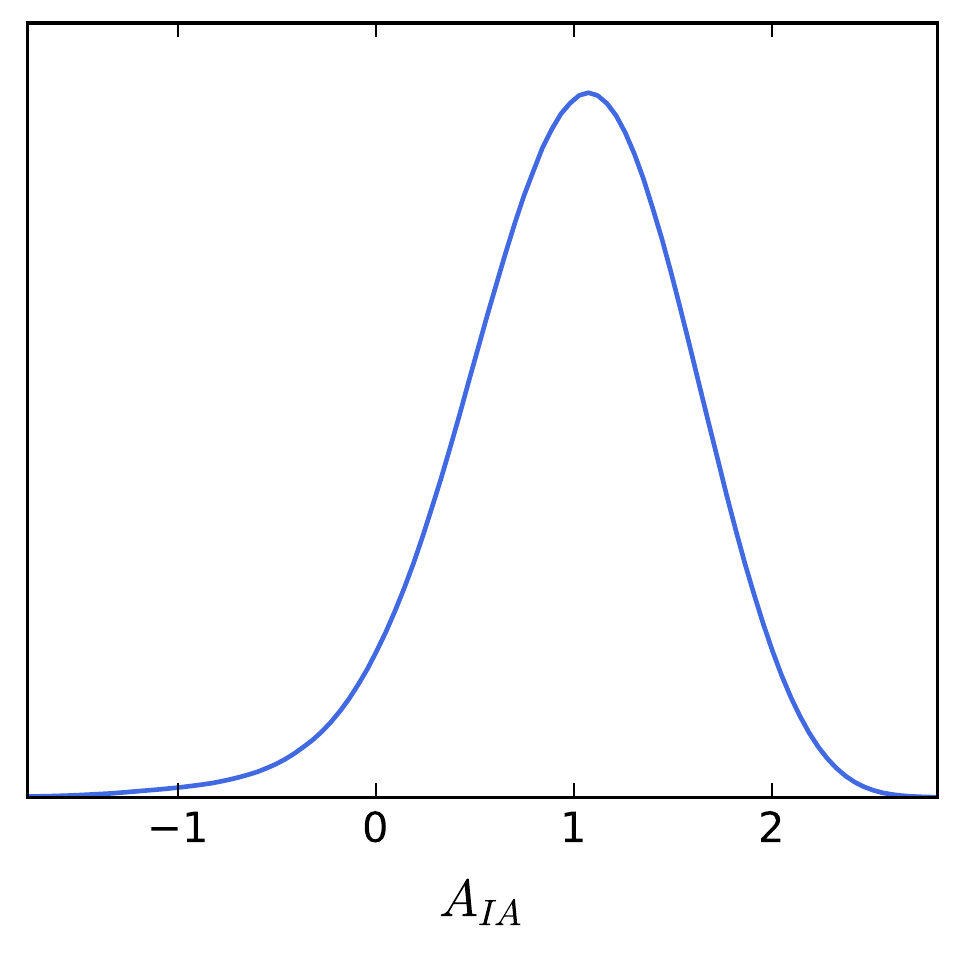}
      \caption{Marginalized posterior distributions of the baryon feedback amplitude $B$ and the intrinsic alignment amplitude $A_{IA}$ for Yukawa interaction model by using KiDS data.}
      \label{BA}
\end{figure}

Fig. \ref{fig.1D_plk} shows the 1D marginalized posterior distributions by using KiDS datasets. Due to large band power uncertainties in the weak lensing measurements, the constraints from KiDS data have wider $68\%$ confidence regions compared to that from Planck data. Note that the amplitude of scalar perturbation $A_s$ and the scalar spectral index $n_s$ are mainly constrained by the priors rather than by the KiDS data. We find that the KiDS data presents a preference for larger values of the scalar potential parameter $\lambda$ than the Planck data, and the best fit values of the coupling parameter $r$ becomes positive which is opposite to the results obtained from Planck data. We also show the impact of baryonic feedback $B$ and intrinsic alignment amplitude $A_{IA}$ of the weak lensing analysis. In Fig. \ref{BA}, we can see that the KiDS data do not strongly constrain the baryon feedback amplitude $B$, which indicates that this astrophysical effect is relatively unimportant in our analysis. Future cosmic shear surveys with higher signal-to-noise measurements and finer binning in angle and redshift or crosscorrelations between lensing and baryonic probes may constrain $B$ to a reasonable level \citep{Hildebrandt2017}. In \cite{Joudaki2017}, they found an almost $2\sigma$ preference for a nonzero intrinsic alignment amplitude in the $\Lambda$CDM model, where $-0.45 < A_{IA} < 2.3$, which is similar to the constraint of $-0.24 < A_{IA} < 2.5$ when considering the fiducial treatment of the systematic uncertainties. As shown in Fig. \ref{BA}, the constraint on the intrinsic alignment amplitude in the Yukawa interaction model is only marginally affected by the extra two free parameters $r$ and $\lambda$ in our analysis, where $-0.15<A_{IA}<2.1(95\%\text{CL})$. We will consider the combination of Planck15+RSD+KiDS below, where this constraint can improve to  $0.65<A_{IA}<1.96(95\%\text{CL})$.

\begin{figure}
 \includegraphics[scale=0.3]{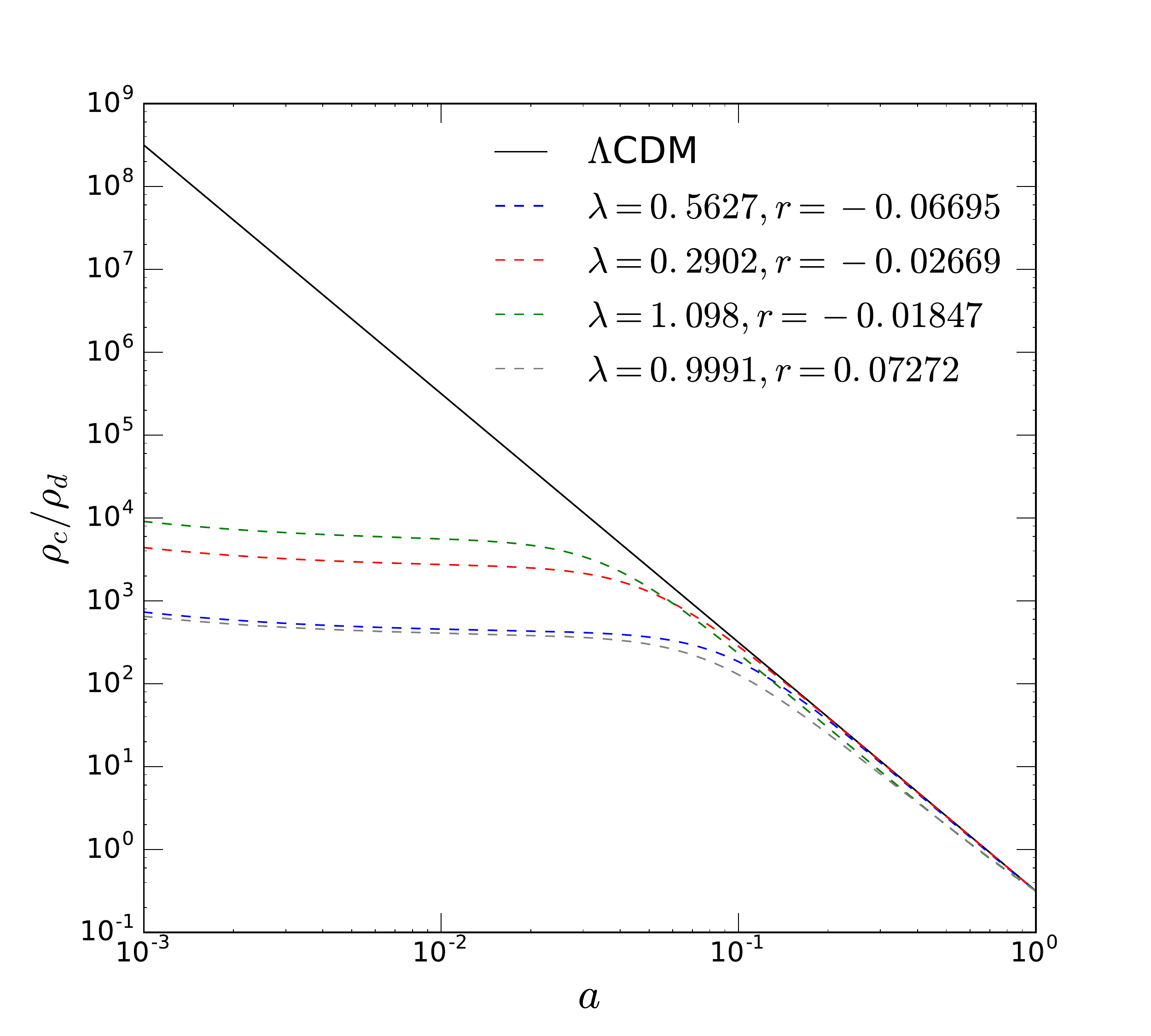}
\caption{Time evolution of the ratio between the energy densities of dark matter and dark energy. The dashed lines correspond to the Yukawa interaction model with different best fit values of $\lambda$ and $r$ listed in Table \ref{tab.pk}, where the blue line corresponds to Planck13, the red line corresponds to Planck15, the green line corresponds to Planck15(+lensing) and the gray one corresponds to KiDS.}
\label{fig.ratio}
\end{figure}

Besides, we find that the best fit values we obtained for $\lambda$ and $r$ in Table \ref{tab.pk} can help to alleviate the coincidence problem. As shown in Fig. \ref{fig.ratio}, we present the time evolution of the ratio between the energy densities of dark matter and dark energy, we can see that the energy densities of dark matter and dark energy in the Yukawa interaction models have more time to be comparable in the past. 

\begin{figure}
 \includegraphics[scale=0.6]{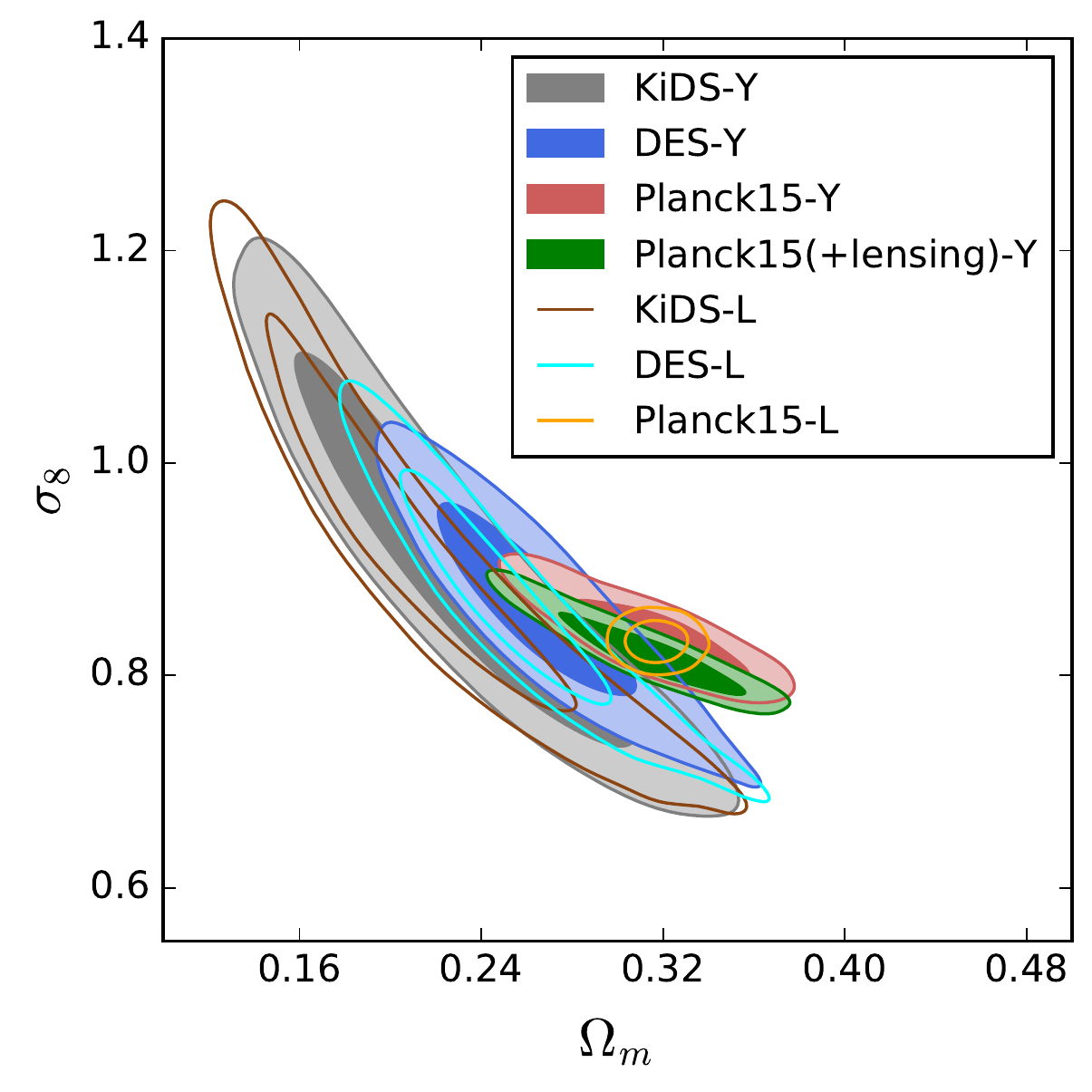}
\caption{Marginalized confidence contours in the  $\sigma_8-\Omega_m$ plane for the $\Lambda$CDM model ($\Lambda$) and Yukawa interaction model (Y). $68\%$ and 95\% confidence levels are shown as inner and outer regions.}
\label{fig.omsi8}
\end{figure}

For the purposes of model selection, we use the deviance information criterion (DIC) \citep{Spiegelhalter2002,An2018} to investigate whether the Yukawa interaction model is more favored by Planck15 and KiDS datasets,  as compared to the $\Lambda$CDM model. DIC is composed of the sum of goodness of fit of a given model and its Bayesian complexity, which is defined as
\begin{equation}
\label{eq.DIC}
{\rm DIC}=\chi_{\rm eff}^2(\bm{\hat\theta})+2p_{\text{D}},
\end{equation}
where $\chi_{\rm{eff}}^2(\bm{\hat\theta})=-2\text{ln}\mathcal{L}_{\rm max}$ is the best-fit effective $\chi^2$ and $\bm{\hat\theta}$ is the parameter vector at the maximum likelihood point. The second term in equation (\ref{eq.DIC}) is the Bayesian complexity expressed as $p_{\text{D}}=\langle\chi_{\rm eff}^2(\bm{\theta})\rangle-\chi_{\rm eff}^2(\bm{\hat\theta})$, where $\langle\chi_{\rm eff}^2(\bm{\theta})\rangle$ represents the mean $\chi^2$ averaged over the posterior distribution. A difference of 10 in DIC between two models constitutes strong preference in favor of the model with the lower DIC estimate, and a difference of 5 in DIC between two models constitutes moderate preference in favor of the model with the lower DIC estimate. When the difference is close to zero, it means one model is not favored over the other \citep{Joudaki2017}. In comparing the Yukawa interaction model with the $\Lambda$CDM model, we take negative values of $\Delta$DIC to indicate a preference in favor of the Yukawa interaction model. We find that Planck15 datasets moderately favors the Yukawa interaction model with a negative value $\Delta\text{DIC}=-5.152$, while the KiDS datasets does not show any preference for the Yukawa interaction model due to the small positive value $\Delta\text{DIC}=0.395$.

In this work we also aim to investigate whether the Yukawa interaction model can alleviate the tension between KiDS and Planck that has been reported for the $\Lambda$CDM model. Fig. \ref{fig.omsi8} shows the parameter constraints in the $\sigma_8$--$\Omega_m$ plane for the Yukawa interaction model and the standard $\Lambda$CDM model which has been presented in Figure 1 of \cite{An2018}. We find that different from the $\Lambda$CDM model, the KiDS and Planck constraint contours of the Yukawa interaction model start to overlap with each other. In order to quantify how much the tension between Planck and KiDS has been reduced by Yukawa interaction, we employ the tension parameter diagnostic. Since current lensing data mainly constrain the $S_8=\sigma_8\sqrt{\Omega_m}$ parameter combination well, the tension parameter $T$ can be defined as~\citep{Joudaki2017}
\begin{equation}
\label{eq.TS8}
T(S_8)=\frac{|\langle S_8^{\text{K}}\rangle-\langle S_8^{\text{P}}\rangle|}{\sqrt{\sigma^2(S_8^{\text{K}})+\sigma^2(S_8^{\text{P}})}},
\end{equation}
where $\langle S_8\rangle$ is the mean value over the posterior distribution and $\sigma$ refers to the symmetric $68\%$ confidence interval about the mean. The superscripts K denotes the KiDS data and P denotes the Planck data. The tension between KiDS and Planck15 datasets for the $\Lambda$CDM and Yukawa interaction models is $2.11\sigma$ and $1.54\sigma$, respectively. If we consider CMB lensing, the discordance between KiDS and Planck15(+lensing) for the Yukawa interaction model can reduce to $1.34\sigma$. From this $T$-parameter test, we find the Yukawa interaction model can moderately alleviate the tension between these two datasets.

As shown in Fig. \ref{fig.omsi8}, we also examine the discordance between DES and Planck. We find that the DES constraint is consistent with that of KiDS in the $\Lambda$CDM model, and the consistency is unchanged in the Yukawa interaction model. Tension between DES and Planck15 datasets for the $\Lambda$CDM model is $1.89\sigma$, and it will reduce to $1.30\sigma$ in the Yukawa interaction model. If we consider CMB lensing, the tension between DES and Planck15(+lensing) for the Yukawa interaction model will be $0.86\sigma$. These indicate that the Yukawa interaction model can also alleviate the discordance between DES and Planck inferred from the $\Lambda$CDM model.

\begin{table*}
\centering
\caption{Best fit values and $68\%$ confidence levels for the cosmological parameters}
\label{tab.pf}
\begin{tabular}{ccccccc}
\hline
    & \multicolumn{2}{c}{Planck15} & \multicolumn{2}{c}{Planck15+RSD} & \multicolumn{2}{c}{Planck15+RSD+KiDS}\\
    \cline{2-7}
    \ \ Parameter\ \  & \ \ \ Best fit\ \ \  & 68\% limits &\ \ \ Best fit\ \ \  & 68\% limits &\ \ \ Best fit\ \ \  & 68\% limits \\
    \hline
$\Omega_bh^2$ & 0.02203 & $0.0222^{+0.000157}_{-0.000157}$ & 0.02229 & $0.02229^{+0.000172}_{-0.000173}$ & 0.02232 & $0.02237^{+0.000153}_{-0.000151}$\\
$\Omega_ch^2$ & 0.1201 & $0.1181^{+0.00305}_{-0.00157}$ & 0.1174 & $0.1167^{+0.00356}_{-0.00156}$ & 0.1108 & $0.1162^{+0.00225}_{-0.00127}$\\
$100\theta_{MC}$ & 1.041 & $1.041^{+0.000336}_{-0.000338}$ & 1.041 & $1.041^{+0.000356}_{-0.000357}$ & 1.041 & $1.041^{+0.000318}_{-0.000316}$\\
$\tau$ & 0.0814 & $0.08287^{+0.0172}_{-0.017}$ & 0.09015 & $0.08137^{+0.0273}_{-0.0249}$ & 0.09088 & $0.08108^{+0.0168}_{-0.0169}$\\
$n_s$ & 0.9632 & $0.9644^{+0.00488}_{-0.00485}$ & 0.9698 & $0.9659^{+0.00554}_{-0.00549}$ & 0.9707 & $0.9694^{+0.00456}_{-0.00454}$\\
${\rm{ln}}(10^{10}A_s)$ & 3.1 & $3.102^{+0.0334}_{-0.0331}$ & 3.107 & $3.095^{+0.0519}_{-0.0467}$ & 3.116 & $3.082^{+0.0319}_{-0.032}$\\
$\lambda$ & 0.2902 & $0.6858^{+0.231}_{-0.676}$ & 1.395 & $1.281^{+0.219}_{-0.0241}$ & 1.2966 & $1.122^{+0.378}_{-0.15}$\\
$r$ & -0.02669 & $-0.01074^{+0.0424}_{-0.0426}$ & 0.01172 & $-0.005202^{+0.0446}_{-0.0452}$ & 0.02748 & $-0.001688^{+0.0353}_{-0.0343}$\\
\hline
$\Omega_\Lambda$ & 0.6827 & $0.6835^{+0.0224}_{-0.0268}$ & 0.644 & $0.6632^{+0.0156}_{-0.0262}$ & 0.7392 & $0.6737^{+0.0165}_{-0.0195}$\\
$\Omega_m$ & 0.3173 & $0.3165^{+0.0268}_{-0.0224}$ & 0.356 & $0.3368^{+0.0262}_{-0.0156}$ & 0.2608 & $0.3263^{+0.0195}_{-0.0165}$\\
$z_{\rm{re}}$ & 10.37 & $10.36^{+1.62}_{-1.37}$ & 11.03 & $10.09^{+2.63}_{-1.9}$ & 11.23 & $9.705^{+1.58}_{-1.41}$\\
$H_0$ & 67.09 & $66.89^{+2.09}_{-2.3}$ & 62.78 & $64.47^{+0.988}_{-2.25}$ & 71.61 & $65.39^{+1.43}_{-1.86}$\\
${\rm{Age}}/{\rm{Gyr}}$ & 13.81 & $13.76^{+0.112}_{-0.037}$ & 13.87 & $13.79^{+0.105}_{-0.0323}$ & 13.57 & $13.79^{+0.0732}_{-0.0359}$\\
\hline
\end{tabular}
\end{table*}

\begin{figure*}
 \includegraphics[scale=0.36]{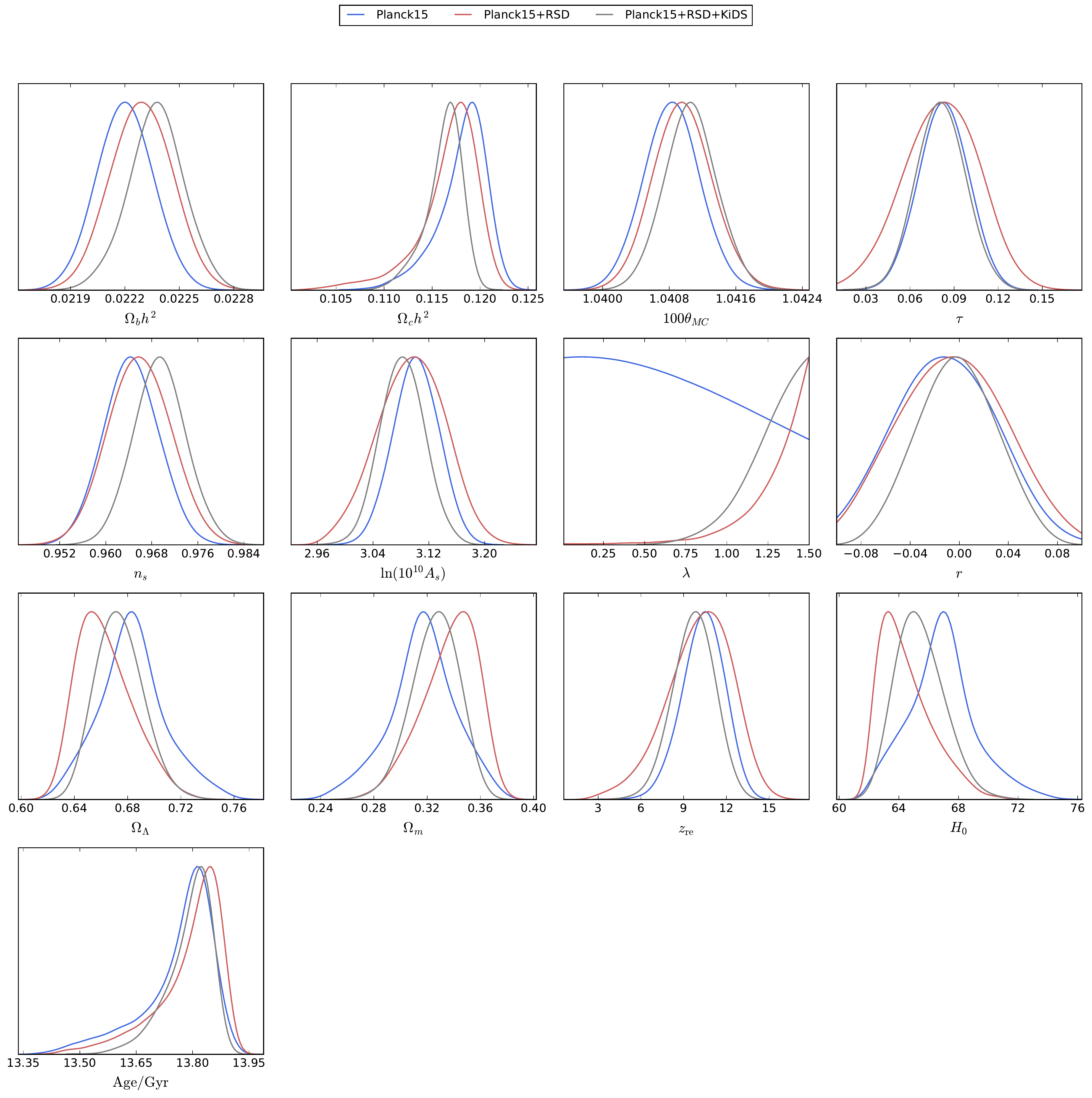}
\caption{1D distributions for the cosmological parameters using Planck, RSD and KiDS data.}
\label{fig.1D_pf}
\end{figure*}

\begin{figure*}
 \includegraphics[scale=0.5]{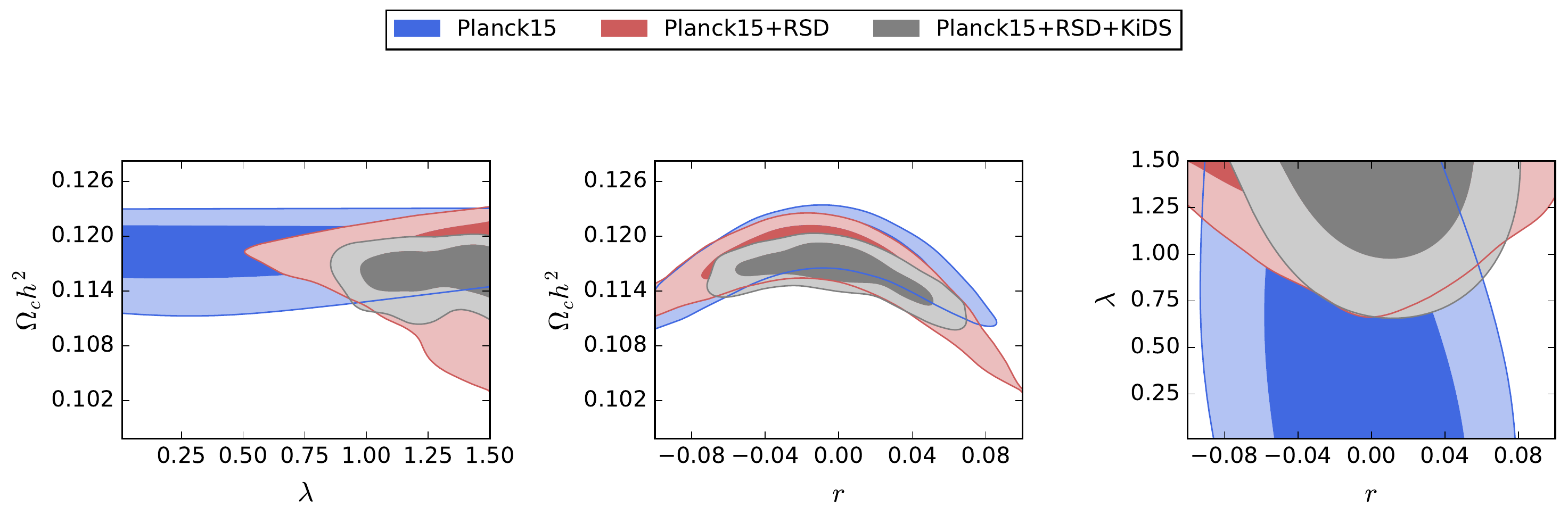}
\caption{2D distributions for selected parameters using Planck, RSD and KiDS data.}
\label{fig.2D_pf}
\end{figure*}

In the following discussion, we add the redshift-space distortions data from large scale structure observations to constrain the Yukawa interaction model. In Table \ref{tab.pf}, we present the best fit values and $68\%$ limits from the joint analysis of Planck15+RSD and Planck15+RSD+KiDS. The 1D and 2D posterior distributions are shown in Fig. \ref{fig.1D_pf} and Fig. \ref{fig.2D_pf}, respectively. In the discussion above we found that the Planck15 data alone can not constrain the scalar potential parameter $\lambda$, while including the RSD data we see that $\lambda$ can be tight constrained, and the RSD data prefers higher values of $\lambda$. For the coupling parameter $r$, Planck15 data alone puts a preference for a negative value, while the joint analysis by combing RSD data together prefers a very small positive interaction term for the best fitting value. These results are compatible with the previous analysis in Section III. We also find that the joint analysis including KiDS data can improve the constraint for the coupling parameter $r$ and prefer a larger positive best fitting value. For the scalar potential parameter $\lambda$, the combination of these three datasets still presents a preference for larger value.

From Fig. \ref{fig.1D_p}, \ref{fig.1D_plk} and \ref{fig.1D_pf}, we observe that the cosmological observations we employed in this work can not break the symmetry around zero value of the coupling parameter $r$. This is due to the behavior of spectrum manifesting some symmetry around the axis of $r=0$ \citep{Costa2014}. In the following section, we will go beyond the linear perturbation study and employ the N-body simulations to test this model in the non-linear structure formation and find whether the nonlinear perturbation can provide richer physics to break down the degeneracy between positive and negative $r$.

\section{Simulations}

We perform N-body simulations using ME-Gadget and the related simulation pipeline \citep{gadget2,2lptic,liao2018ccvt,me-gadget-prd,me-gadget-apjl} to investigate the non-linear structure formation in the Yukawa-type dark matter and dark energy interaction model. ME-Gadget is a modified version of publicly available Gadget2 code. We can use ME-Gadget to perform N-body simulations for Interacting Dark Energy model \citep{me-gadget-prd}. The Yukawa interaction model is one kind of Interacting Dark Energy model, ME-Gadget can also solve the non-linear structure formation accurately in the Yukawa interaction model. In order to take Yukawa-type dark matter and dark energy into account in the N-body simulation, we need to modify both the initial condition for the simulation and the simulation procedure. We use the modified CAMB to generate the linear matter power spectrum at $z=49$, and use a modified version of 2LPTic \citep{2lptic} to load in the CAMB generated matter power spectrum and generate the initial condition for the simulation. In the simulation procedure, four major difference from $\Lambda$CDM simulation need to be taken into account: a) the expansion rate of the universe shall be changed accordingly; b) $\rho_c\propto a^{-3}$ is no longer true, where $a$ is the scale factor, therefore, the particle mass in the simulation is varying; c) the particles in the simulation receive additional drag force $\dot{v}=\alpha(a)v$ due to the interaction with the background dark energy field, just like a car passes through the wind and feels friction; d) additional gravitational force caused by the perturbation of dark energy field, which can be treated as an effective gravitational constant. For more details, please refer to \cite{me-gadget-prd}. We calculate the Hubble expansion rate $H(z)/H_0$, the particle mass variation $m(a)$, the drag force parameter $\alpha(a)$ and the change of effective gravitational constant $G_{eff}(a)/G$. We take these four effects into account using ME-Gadget and perform N-body simulations.

\begin{figure}
\includegraphics[width=0.45\textwidth]{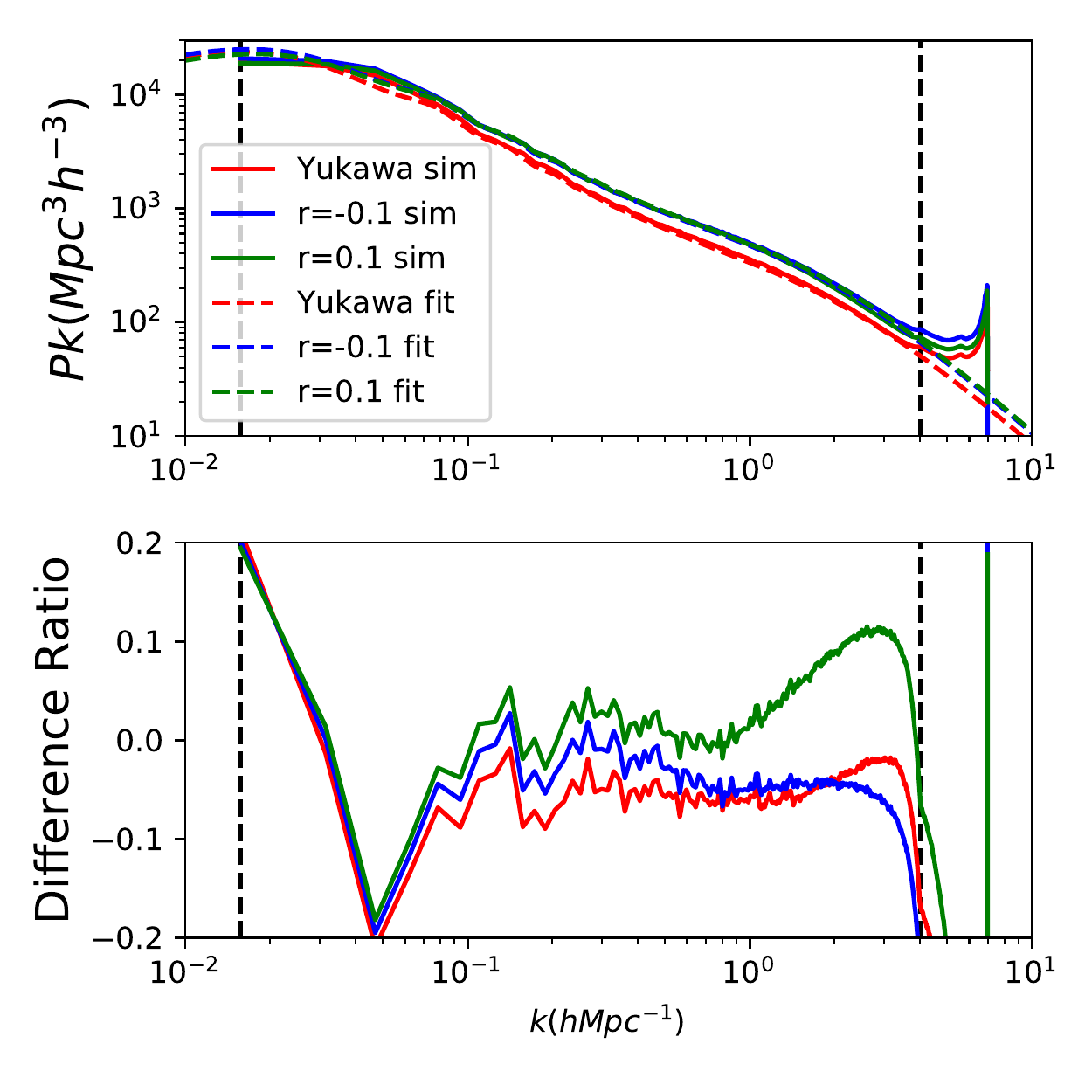}
\caption{Upper panel: The measured matter power spectrum of three simulations are shown in solid lines and the matter power spectrum calculated by halofit are shown in dashed lines. The black vertical line on the left (right) shows the boxsize limit (Nyquist limit). Lower panel: The ratio of difference between the simulation and halofit is given in different colors. Despite the large difference on the left due to cosmic variance, it is clear that halofit can provide accurate ($<10\%$ difference) estimation for the matter power spectrum.}\label{fig.pk}
\end{figure}

Since we have used KiDS weak gravitational lensing data to set constraints, it is important to notice that in fact the non-linear correction of matter power spectrum provided in CAMB, so-called halofit, is not consistent with the Yukawa interaction model. We use the simulations to confirm that, though halofit is not consistent with the Yukawa interaction model, it can still provide reasonably accurate matter power spectrum. Three N-body simulations with Yukawa interaction, called Yukawa, $r=-0.1$ and $r=0.1$, and one $\Lambda$CDM simulation are performed in this work. Their  parameters are given in Table~\ref{tab.sim}, where NumPart is short for number of particles, SoftLen is short for softening length, and the cosmological parameters in Yukawa case are taken from Planck15 constraint results. We use the same random seed to generate the initial conditions for the simulations and measure the matter power spectrum with ComputePk code \citep{computepk}. The results are shown in Fig.~\ref{fig.pk}. We found that the difference of matter power spectrum between simulations and halofit is less than $10\%$, which is accurate enough for our calculation of KiDS weak lensing constraints.

\begin{table}
\caption{Parameters for N-body simulations.}
\label{tab.sim}
\begin{tabular}{ccccc}
\hline
Parameter & $\Lambda$CDM & Yukawa & r=-0.1 & r=0.1 \\
\hline
$\Omega_bh^2$ & 0.02225 & 0.0222 & 0.02225 & 0.02225 \\
$\Omega_ch^2$ & 0.1198 & 0.1181 & 0.1198 & 0.1198 \\
$100\theta_{MC}$ & 1.04077 & 1.041 & 1.04077 & 1.04077 \\
$\tau$ & 0.079 & 0.08287 & 0.079 & 0.079 \\
$n_s$ & 0.9645 & 0.9644 & 0.9645 & 0.9645 \\
${\rm{ln}}(10^{10})A_s$ & 3.094 & 3.102 & 3.094 & 3.094 \\
$\lambda$ & -- & 0.6858 & 1.2247 & 1.2247 \\
$r$ & -- & -0.01074 & -0.1 & 0.1 \\
$H_0$ & 67.27 & 66.89 & 67.27 & 67.27 \\
Boxsize/$h^{-1}$Mpc & 400 & 400 & 400 & 400 \\
NumPart & $512^3$ & $512^3$ & $512^3$ & $512^3$ \\
SoftLen/$h^{-1}$kpc & 25 & 25 & 25 & 25 \\ 
\hline
\end{tabular}
\end{table}

\begin{figure}
\includegraphics[width=0.45\textwidth]{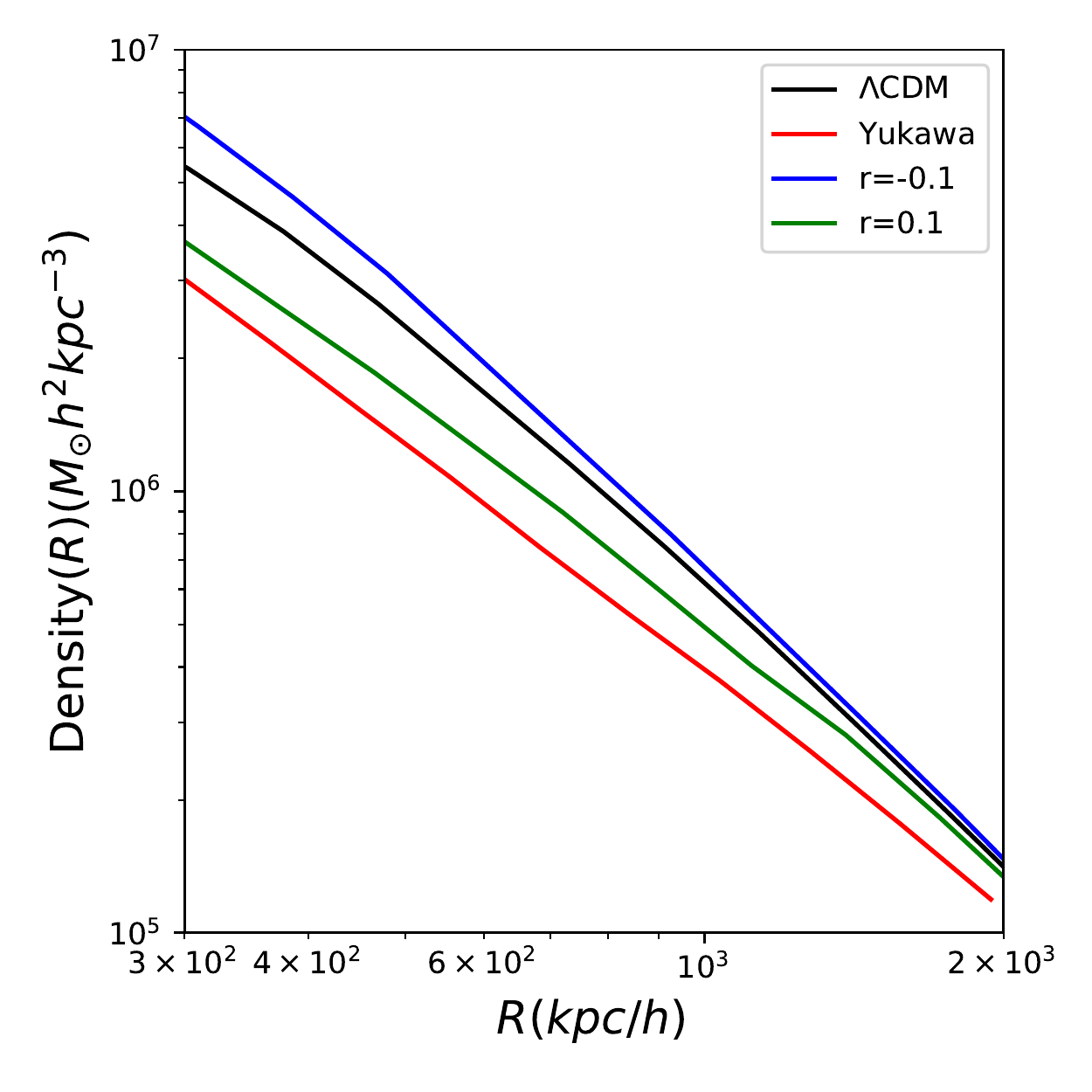}
\caption{The halo density profile of the most massive halo in the simulations are shown. We have checked that these four halos are correnspondant, which means they share similar environment. Though the halo density profiles of $r=-0.1$ and $r=0.1$ are similar at $R\sim 2000$ kpc/h, large difference can be found in the inner region of the halos.}\label{fig.hdp}
\end{figure}

Using N-body simulations, we also confirm that the effects of $r=-0.1$ and $r=0.1$ on the matter power spectrum are very similar. We investigate the halo density profile of the most massive halos in the simulations in Fig.~\ref{fig.hdp}. The halos are identified by Amiga's Halo Finder \citep{AHF}. The red line is lower than the other three, because it takes a lower value of $\Omega_ch^2$. Comparing to $\Lambda$CDM, $r=-0.1$ and $r=0.1$ depart to different direction from the edge of the halo to the center. The halo in $r=-0.1$ simulation is more concentrate than the $\Lambda$CDM case, while in $r=0.1$, the halo is looser. The inter region of halos is highly non-linear, and the difference is large between $r=-0.1$ and $r=0.1$. \cite{me-gadget-apjl} introduces that it is possible to use galaxy-galaxy lensing to distinguish different interacting dark energy models by investigating the structure of the dark matter halos.  This large difference of halo density profile indicates that we can use galaxy-galaxy lensing to break the degeneracy between positive and negative $r$. The detail discussion is beyond the scope of the present work and will be fully discussed in the future.

\section{Conclusions}

In this work we have obtained the observational constraints on the Yukawa-type dark matter and dark energy interaction model using both weak gravitational lensing data from the KiDS and the updated CMB data from Planck. We find that the constraints from Planck15 have been clearly improved compared with that from Planck13. Due to large band power uncertainties, the KiDS datasets alone  have less constraining power than the Planck datasets. From the constraint results we find that the Planck data induces evidence for a negative value of the coupling parameter $r$, while the KiDS data presents a preference for a positive value of $r$. These two datasets can not constrain the scalar potential parameter $\lambda$. So we add new complementary datasets from large scale structure observations, RSD, to investigate the constraints on the Yukawa interaction model. We find that RSD data prefers a very small positive value for $r$ and can put a tight constraint on $\lambda$ with higher values. Besides we find that the interaction in the Yukawa model can help to alleviate the coincidence problem, accommodating longer period for dark matter and dark energy comparable to each other.

We also investigate whether the Yukawa interaction model can alleviate the $\Lambda$CDM  discordance problem between KiDS and Planck datasets, and whether the Yukawa interaction model is favored by these datasets as compared to the $\Lambda$CDM model. Employing the DIC and tension parameter diagnostics, the Yukawa interaction model is found to be moderately favored by the Planck datasets, and able to alleviate the tension between KiDS and Planck. We also find that the tension between DES and Planck datasets inferred from the $\Lambda$CDM model can be reduced by the Yukawa interaction model. With the improvement of the weak lensing measurements, the desired concordance between weak lensing and CMB datasets can be used to support the interaction between dark sectors. 

In July 2018, Planck Collaboration has released their final results on constraining the cosmological parameters \citep{Planck2018}. We would like to use the updated likelihood and data to test our model, once they are available. We expect that future more precise Planck 2018 data can help us to draw more accurate conclusions on examining the Yukawa interaction model. Compared to Planck 2015 results, the improved measurements of large-scale polarization allow the reionization optical depth to be measured with higher precision, this in turn will affect other correlated parameters , and the improved modelling of the small-scale polarization will also lead to more robust constraints on the cosmological parameters \citep{Planck2018}.

By performing self-consistent N-body simulations, we find that the assumed halofit model can accurately estimate the non-linear matter power spectrum for Yukawa interaction model. Thus, the methodology we use to set constraints from KiDS cosmic shear measurements is correct. By measuring the halo density profile, we find that the difference between $r=-0.1$ and $r=0.1$ is significant. Hopefully we may break the symmetry around zero value for the coupling parameter $r$ using galaxy-galaxy lensing. We will study the non-linear structure formation in Yukawa interaction model in more detail and put further constraints on this model from galaxy-galaxy lensing in the future.

\section*{Acknowledgements}

This work was partially supported by National Science Foundation of China (No.11835009). A. C. acknowledges FAPESP and CAPES for the financial support under grant number 2013/26496-2, S\~ao Paulo Research Foundation (FAPESP). J.Z acknowledges the support from China Postdoctoral Science Foundation 2018M632097. We also acknowledge the discussions in HOUYI workshop.

\bibliographystyle{mnras}
\bibliography{Yukawa} 
\bsp
\label{lastpage}
\end{document}